\documentclass[twocolumn]{aastex701}
\usepackage{xcolor}
\usepackage{gensymb}
\usepackage{amsmath}
\usepackage{orcidlink}
\usepackage{graphicx}
\usepackage{subcaption}
\usepackage{hyperref}
\usepackage{placeins}

\makeatletter
\newcommand*{\rom}[1]{\expandafter\@slowromancap\romannumeral #1@}
\makeatother

\newcommand{\e}[1]{\times 10^{#1}}
\newcommand{\hatcurhtr}{TIC183374187}                                  
\newcommand{\hatcurCCra}{\ensuremath{23^{\mathrm h}50^{\mathrm m}27.849{\mathrm s}}} 
\newcommand{\hatcurCCdec}{\ensuremath{-41{\arcdeg}24{\arcmin}44.72{\arcsec}}}
\newcommand{\hatcurCCgaiamG}{\ensuremath{14.50510\pm0.00030}}          
\newcommand{\hatcurCCgaiamBP}{\ensuremath{14.8233\pm0.0013}}           
\newcommand{\hatcurCCgaiamRP}{\ensuremath{14.02110\pm0.00080}}         
\newcommand{\hatcurCCtwomassJmag}{\ensuremath{13.455\pm0.022}}         
\newcommand{\hatcurCCtwomassHmag}{\ensuremath{13.159\pm0.027}}         
\newcommand{\hatcurCCtwomassKmag}{\ensuremath{13.134\pm0.029}}         
\newcommand{\hatcurCCWonemag}{\ensuremath{13.078\pm0.023}}             
\newcommand{\hatcurCCWtwomag}{\ensuremath{13.093\pm0.027}}             
\newcommand{\hatcurLCrprstar}{\ensuremath{0.0959\pm0.0041}}            
\newcommand{\hatcurLCimp}{\ensuremath{0.480_{-0.106}^{+0.080}}}        
\newcommand{\hatcurLCdur}{\ensuremath{0.1813\pm0.0066}}                
\newcommand{\hatcurLCP}{\ensuremath{5.05907\pm0.00059}}                
\newcommand{\hatcurLCT}{\ensuremath{1367.2682\pm0.0032}}               
\newcommand{\hatcurISOm}{\ensuremath{0.853\pm0.048}}                   
\newcommand{\hatcurISOr}{\ensuremath{1.344\pm0.043}}                   
\newcommand{\hatcurISOrho}{\ensuremath{0.495\pm0.040}}                 
\newcommand{\hatcurISOlogg}{\ensuremath{4.112\pm0.025}}                
\newcommand{\hatcurISOteff}{\ensuremath{5768\pm58}}                    
\newcommand{\hatcurISOzfeh}{\ensuremath{-0.39\pm0.10}}                 
\newcommand{\hatcurISOage}{\ensuremath{12.31_{-0.64}^{+1.29}}}         
\newcommand{\hatcurRVK}{\ensuremath{74\pm25}}                          
\newcommand{\hatcurRVgamma}{\ensuremath{33\pm17}}                      
\newcommand{\hatcurPPlogg}{\ensuremath{2.95_{-0.16}^{+0.12}}}          
\newcommand{\hatcurPPar}{\ensuremath{8.75\pm0.23}}                     
\newcommand{\hatcurPPrho}{\ensuremath{0.35\pm0.14}}                    
\newcommand{\hatcurPPm}{\ensuremath{0.56\pm0.19}}                      
\newcommand{\hatcurPPr}{\ensuremath{1.252\pm0.070}}                    
\newcommand{\hatcurPPteff}{\ensuremath{1379\pm23}}                     
\newcommand{\hatcurPPtheta}{\ensuremath{0.057\pm0.020}}                
\newcommand{\hatcurPPtcirc}{\ensuremath{550\pm260}}                    
\newcommand{\hatcurPPtinfall}{\ensuremath{3770_{-950}^{+1720}}}        
\newcommand{\hatcurXAv}{\ensuremath{0.0270_{-0.0120}^{+0.0070}}}       
\newcommand{\hatcurXdistred}{\ensuremath{1218\pm30}}                   

\begin{document}
\linenumbers

\title{The T16 Planet Hunt: 10,000 New Planet Candidates from TESS Cycle 1 and the Confirmation of a Hot Jupiter Around TIC 183374187\footnote{This paper includes data gathered with the 6.5 meter Magellan Telescopes located at Las Campanas Observatory, Chile. }}

\correspondingauthor{Joshua T. Roth}
\email{joshua.roth@princeton.edu}

\author[0009-0002-4817-5304]{Joshua T. Roth}
\email{joshua.roth@princeton.edu}
\affiliation{Department of Astrophysical Sciences, Princeton University, 4 Ivy Lane, Princeton, NJ 08544, USA}

\author[0000-0001-8732-6166]{Joel D. Hartman}
\email{jhartman@astro.princeton.edu}
\affiliation{Department of Astrophysical Sciences, Princeton University, 4 Ivy Lane, Princeton, NJ 08544, USA}

\author[0000-0001-7204-6727]{Gáspár Á. Bakos}
\email{gbakos@astro.princeton.edu}
\affiliation{Department of Astrophysical Sciences, Princeton University, 4 Ivy Lane, Princeton, NJ 08544, USA}

\author[0000-0001-7961-3907]{Samuel W. Yee}
\email{syee@astro.ucla.edu}
\affiliation{Department of Physics \& Astronomy, University of California Los Angeles, Los Angeles, CA 90095, USA}

\author[0000-0002-0514-5538]{Luke G. Bouma}
\email{lbouma@carnegiescience.edu}
\affiliation{Carnegie Science Observatories, 813 Santa Barbara Street, Pasadena, CA 91101, USA}

\author[0000-0001-9261-8366]{Jhon Yana Galarza}
\email{jyanagalarza@carnegiescience.edu}
\affiliation{Departamento de Astronomía, Facultad de Ciencias Físicas y Matemáticas Universidad de Concepción, Av. Esteban Iturra s/n Barrio Universitario, Casilla 160-C, Chile}
\affiliation{Carnegie Science Observatories, 813 Santa Barbara Street, Pasadena, CA 91101, USA}

\author[0009-0008-2801-5040]{Johanna K. Teske}
\email{jteske@carnegiescience.edu}
\affiliation{Carnegie Science Observatories, 813 Santa Barbara Street, Pasadena, CA 91101, USA}
\affiliation{Earth and Planets Laboratory, Carnegie Institution for Science, 5241 Broad Branch Rd NW, Washington, DC 20015}

\author[0000-0003-1305-3761]{R.P. Butler}
\email{pbutler@carnegiescience.edu}
\affiliation{Earth and Planets Laboratory, Carnegie Institution for Science, 5241 Broad Branch Rd NW, Washington, DC 20015}

\author[0000-0002-5226-787X]{Jeffrey D. Crane}
\email{crane@carnegiescience.edu}
\affiliation{Carnegie Science Observatories, 813 Santa Barbara Street, Pasadena, CA 91101, USA}

\author[0000-0002-8681-6136]{Steve Shectman}
\email{shec@carnegiescience.edu}
\affiliation{Carnegie Science Observatories, 813 Santa Barbara Street, Pasadena, CA 91101, USA}

\author[0000-0003-0412-9664]{David Osip}
\email{dosip@carnegiescience.edu}
\affiliation{Las Campanas Observatory, Carnegie Institution for Science, Colina el Pino, Casilla 601 La Serena, Chile}

\author[0000-0003-2527-1475]{Shreyas Vissapragada}
\email{svissapragada@carnegiescience.edu}
\affiliation{Carnegie Science Observatories, 813 Santa Barbara Street, Pasadena, CA 91101, USA}

\author{Yuri Beletsky}
\email{ybeletsky@carnegiescience.edu}
\affiliation{Earth and Planets Laboratory, Carnegie Institution for Science, 5241 Broad Branch Rd NW, Washington, DC 20015}

\author[0000-0001-8401-4300]{Shubham Kanodia}
\email{skanodia@carnegiescience.edu}
\affiliation{Earth and Planets Laboratory, Carnegie Institution for Science, 5241 Broad Branch Rd NW, Washington, DC 20015}

\author[0000-0002-2036-2311]{Yadira Gaibor}
\email{ygaibor@mit.edu}
\affiliation{Department of Physics, Massachusetts Institute of Technology, 77 Massachusetts Avenue, Cambridge, MA 02139, USA}
\affiliation{Kavli Institute for Astrophysics and Space Research, Massachusetts Institute of Technology, Cambridge, MA 02139, USA}

\begin{abstract}
The T$16$ project has produced a uniformly detrended and systematics-corrected set of $83,717,159$ TESS Cycle $1$ full-frame image light curves for stars observed by TESS in its primary mission down to $T=16$\,mag, enabling sensitive transit searches beyond the official TESS pipelines. While most existing TESS planet searches focus on relatively bright targets, planet occurrence rates suggest that a substantial number of planets should exist around fainter stars. We therefore use the T$16$ light curves to conduct a semi-automated search for transiting exoplanets across the full Cycle $1$ FFI sample, resulting in $11,554$ planet candidates orbiting stars down to 16th magnitude in the TESS band with orbital periods between $0.5$ and $27$ days. Of these, $10,091$ are new planet candidates, and
%
%
$411$ are single-transit events, for which we do not attempt to determine orbital parameters. The remaining $1,052$ candidates are previously known TESS candidates. We validate our pipeline through Magellan/PFS radial-velocity follow-up measurements on one of our candidate hosts, TIC $183374187$, a metal poor thick-disk star, confirming the signal as newly identified hot Jupiter. This detection demonstrates our pipeline's ability to identify real, previously undiscovered, transiting planets. Overall, this work shows that large-scale, machine-learning–assisted transit searches of TESS full-frame images can significantly expand the census of transiting planet candidates, particularly around faint stars, providing a rich target set for future validation and follow-up efforts. Our findings more than double the number of known TESS exoplanet candidates.
\end{abstract}

\keywords{exoplanets: detection --- methods: data analysis --- techniques: photometric, radial velocity}

\section{Introduction} \label{sec:intro}

Since the first detection of a transiting planet, HD $209458$b, in 1999 \citep{Charbonneau_2000_firstTransit}, the transit method has proven to be the most successful exoplanet detection method to date, accounting for about three quarters of confirmed exoplanets \citep{Exoplanet_archive_Christensen}. Transits furthermore provide the unique opportunity to study the atmospheres of planets via their spectroscopic features. The TESS mission \citep{TESS}, which was launched in April of $2018$ and began operations in that same year, is contributing an ever growing number of planet candidates, thanks to its nearly all-sky coverage and high-precision photometry of hundreds of millions of stars. TESS is currently in its third extended mission, and to date has discovered nearly $8000$ transiting planet candidates, of which $760$ have been confirmed at the time of this writing (March $12^{\mathrm{th}}\,2026$, \citealp{Exoplanet_archive_Christensen}).

Many studies have been conducted with the aim of predicting the planetary yield of the TESS primary mission (e.g. \citealp{Sullivan_2015_TESS_yield, Bouma_2017_TESS_yield, Barclay_2018_TESS_yield, Huang_2018_TESS_yield}). These studies found that $10,000$s of planets could be found around faint TESS targets, much exceeding the current sample of planet candidates around such stars. These large predicted yields arise because transit surveys are fundamentally signal-to-noise limited rather than magnitude limited. For intrinsically large planets such as Hot Jupiters, transits remain detectable out to large distances, greatly expanding the searchable volume. In contrast, most transit searches to date have been effectively magnitude limited in order to facilitate follow-up observations, resulting in a substantial incompleteness as the faint end.
Other studies exploring the expected yield around cool stars and M-dwarfs (e.g. \citealp{Ballard_2019_TESS_M_dwarfs, Murihead_2018_TESS_M_dwarfs}) predict that thousands of planets orbiting M-dwarfs can be found in TESS observations, which greatly exceeds the current sample of about $160$ confirmed TESS discovered planets orbiting M-dwarfs. In general, surveys estimate the occurrence rate of planets around Milky Way stars to be greater than one planet per star, though the occurence rate depends heavily on stellar type and metallicity (e.g. \citealp{OccurenceRate_Cassan, Mulders_2018_OccurenceRate, Zhang_2023_OccurenceRate}, see \citet{Winn_2024_Occurrence} for a useful review). 

These predictions highlight the merit of conducting systematic, deep transit searches in TESS full frame images (FFIs), as we expect to find large numbers of new planet candidates at fainter magnitudes. The recently published T$16$ project \citep{T16_Joel} provides high precision, image subtraction light curves for all stars in TESS FFIs down to TESS magnitude $T<16$, by expanding the image subtraction process used by \citet{CPIDS_Bouma} to all stars, providing around $80,000,000$ light curves for TESS Cycle $1$ alone. This work follows up on the T$16$ light curve generation to search these light curves for transiting planet candidates. While previous TESS transit searches were successful at identifying numerous planets (e.g. \citealp{Huang_2020_QLP,Huang_2020b_QLP,Guerrero_2021_SPOC}), they focused on subgroups of the overall available TESS FFI sample, like stars brighter than $T=13.5$ (e.g. \citealp{Kunimoto_2022_TransitSearch}) or young stars and cluster stars (e.g. CDIPS,  \citealp{CPIDS_Bouma}). This work is poised to expand these searches to all TESS FFI targets brighter than $T=16$, significantly increasing the yield of new planet candidates, while noting that following up large numbers of ultra-faint targets will prove challenging. The first set of $\sim 10,000$ such new planet candidates is presented in this work.

The expansive set of T$16$ light curves provides a unique opportunity to drastically increase our sample of transiting exoplanets, which can be used to address many open questions in exoplanetary science. Due to the geometrical transit probability decreasing drastically with orbital distance and the transit signal depending on the ratio of a planet's radius to its host star's radius, the transit method is inherently biased towards detection of hot Jupiters, large planets orbiting close to their respective host stars. A major open question in the study of hot Jupiters is determining their formation pathways (e.g. \citealp{InSituFormation_Batygin, DiskMigration51Peg_Lin, PlanetPlanet_Scattering_Chatterjee, SecularChaos_Wu}). While hot Jupiters constitute a prominent planetary type in our sample, our transit search also discovers other planetary types like warm Jupiters (giant planets with orbital periods $10 \text{ days} < P <100 \text{ days}$), and Neptunes, even potential sub-Neptunes and super-Earths. These planets can be used to study a variety of features in the exoplanet orbital period-radius space, such as the origin of the Neptune desert, which describes a dearth of Neptunian type planets with orbital periods of less than $3$ days \citep{Maze_2016_Neptune_desert, Castro_2024_Neptunian_desert}, the origin of the sub-Neptunian radius gap, which constitutes a deficit of planets with radii between $1.5-2R_\oplus$ \citep{Fulton_2017_radius_gap}, or the origin of the high occurrence rate of super-Earths (e.g. \citealp{Howard_2010_SuperEarths}).  

This paper is organized as follows. In Section \ref{sec:data} we describe the TESS data and our extracted light curves, in Section \ref{sec:method} we describe our machine learning-based classification routine and we describe our results in Section \ref{sec:results}, including one new hot Jupiter confirmation around TIC $183374187$, described in Section \ref{sec:confirmation}. Section \ref{sec:discussion} shows our discussion of the results, including potential limitations and future work and our conclusion are given in Section \ref{sec:conclusions}. 
\section{Light curve generation} \label{sec:data}

\subsection{TESS Data}
The T$16$ project is continuously producing light curves for more TESS sectors. The complete set for the first year of TESS operations, Cycle 1, was recently published \citep{T16_Joel}, and the follow-up work presented here focuses exclusively on this observing cycle. Cycle 1 was split into $13$ sectors, each covering two consecutive $13.7$ day orbits of the TESS spacecraft. At the end of each orbit, data collection was paused for about three hours to allow downlink to Earth. In addition, the spacecraft performed momentum dumps every $2.5$ days to reset the reaction wheels, which introduced brief pointing instabilities. These interruptions and pointing instabilities can produce systematic signals in the extracted light curves that need to be addressed during detrending and post-processing. Thus, each individual light curve is taken over a baseline of $27$ days. 

The TESS spacecraft possesses four wide field optical cameras, which are aligned to cover a total field of view of $24\degree \times 96\degree$. Each camera has an effective aperture size of $10\,\mathrm{cm}$ and images onto four back-illuminated CCDs from MIT/Lincoln Lab that contain $2048 \times 2048$ pixels used for imaging, resulting in $16$ distinct camera and CCD combinations per sector. In this project we treat each sector, camera, and CCD combination independently, but follow up work will include combining observations from multiple sectors to extend our search to longer orbital periods and smaller transiting planets. Each pixel subtends around 21 arcseconds on each side, which can lead to blending (multiple, unresolved sources contributing to the same signal).  

The light curves used in this process were obtained from the TESS full-frame images (FFIs). The TESS spacecraft takes exposures at a 2-second cadence. These individual images are grouped and summed at a $30$-minute cadence to increase the effective exposure time to create the FFIs. Cosmic ray impacts on the CCDs are mitigated by excluding the highest and lowest pixels, details can be found in \cite{TESS_Instrument_Handbook}. 

\subsection{The T16 project}

There are currently at least twelve efforts producing TESS FFI light curves: The MIT Quick-Look Pipeline project \citep{Huang_2020_QLP}, the TESS Science Processing Operations Center Full Frame Image project (SPOC-FFI, \citealp{SPOC_LCs}), the TESS \textit{Gaia} Light Curve project \citep{TGLC}, the Goddard Space Flight Center eleanor lite project \citep{ELEANOR-LITE}, the DIAMANTE project \citep{DIAMANTE_1,DIAMANTE_2}, the PATHOS project \citep{PATHOS_1, PATHOS_2, PATHOS_3, PATHOS_4}, the Cluster Difference Imaging Photometric Survey (CDIPS, \citealp{CPIDS_Bouma}), the TEQUILA project \citep{Ogunwale_2025_TESS_LCs}, a pipeline focused on TESS transients \citep{Fausnaugh_2021_transientTESSLCs}, the DIA pipeline \citep{Oelkers_2018_DIA}, \texttt{TESSreduce}, which primarily focuses on transients \citep{Ridden-Harper_2021_TESSreduce}, and the T$16$ project, which is the basis of the transit search conducted in this project. Detailed information on the T$16$ project can be found in \cite{T16_Joel}, but for convenience, a summarized description of the data reduction methods is given here. 

\subsubsection{Photometry}

In contrast to most of the other TESS light curve projects listed above, the T$16$ project uses Difference Imaging Photometry to extract the light curves from the TESS FFIs. Difference Imaging Photometry was first introduced by \cite{difference_imaging_tomaney_crotts} in the context of microlensing surveys and proves especially useful in crowded fields, for which other methods like PSF or aperture photometry struggle. An optimized method for image subtraction was introduced by \cite{image_sub_lupton_alard}. In practice, difference imaging photometry works by constructing a high-quality reference frame from a stack of images and subtracting this reference from each subsequent exposure. This way, constant sources and background structure are removed, leaving only the variable flux components. The light curve of a given source can then be reconstructed from these differential flux measurements. 

The Difference Imaging Photometry for the T$16$ project was conducted analogous to CDIPS, for a detailed description the reader is referred to \cite{CPIDS_Bouma}. The first major step is a background estimation using a median box filter followed by Gaussian blurring using a kernel of size 48 pixels. The resulting smooth background model is then subtracted. Pixels exhibiting saturation or other erroneous values are masked, as are entire frames affected by momentum dumps or other instrumental systematics. 

Next, the World Coordinate System (WCS) solution is verified, and the \textit{Gaia} Data Release $2$ catalog \citep{Gaia_DR2} is used as the source catalog. \textit{Gaia} coordinates are projected onto each frame and converted from celestial to pixel coordinates. Here the T$16$ project differs from CDIPS: T$16$ includes all \textit{Gaia} DR2 sources with $T < 16$ mag, where T is calculated by fitting a polynomial to the \textit{Gaia} DR2 photometry, whereas CDIPS only includes all sources with $G_{rp} < 13$, as well as open-cluster members and young stars with $G_{rp} < 16$.

After WCS verification a reference image is produced by aligning around $50$ science frames and matching their flux, background, and point spread function (PSF) using \texttt{ficonv}. This reference image is then convolved to match each individual science frame. Photometry is then performed on stars using \texttt{fiphot}, which is included in the FITSH package \citep{fiphot_pal}. Using the \textit{Gaia} astrometric positions, three circular apertures of radii $1.0$, $1.5$ and $2.25$ pixels are placed on each source, and differential flux photometry is measured using \texttt{fiphot}. The reference flux from each source is determined using \textit{Gaia} DR2 rather than directly measured to avoid issues due to crowding, since TESS has a lower resolution than \textit{Gaia}. This approach mitigates contamination from non-variable but blended neighboring sources that are resolved in \textit{Gaia} in the sense that the reference flux of the source is accurate, and thus the measured flux variations are properly scaled into magnitude variations. However, unresolved neighbors can still introduce contaminated signals, a process we refer to as blending. 

\begin{figure*}
    \centering
    \includegraphics[width=\linewidth]{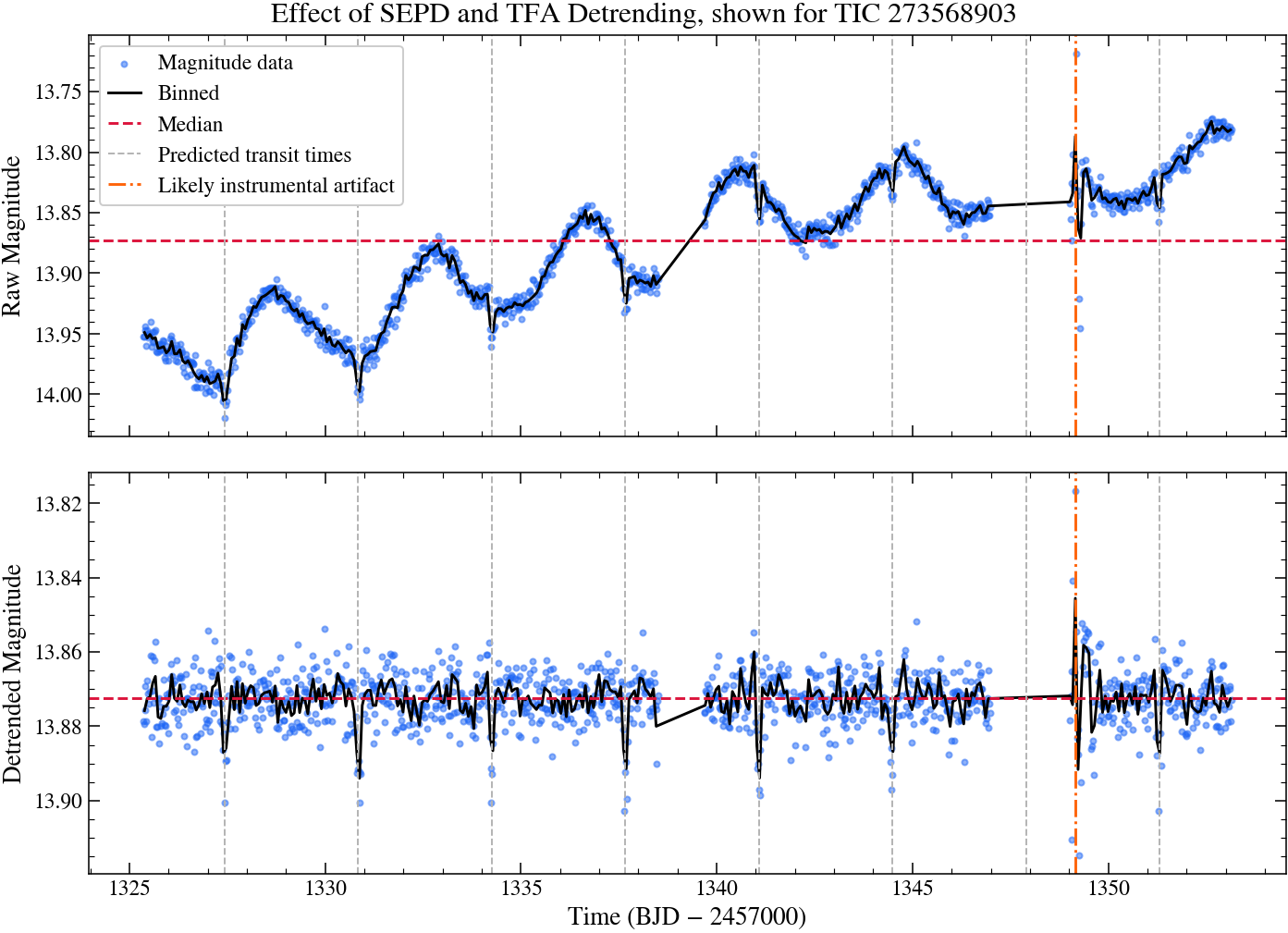}
    \caption{Comparison between the raw (top) and SEPD plus TFA detrended (bottom) light curves for TIC $273568903$. The raw light curve exhibits strong sinusoidal variation as well as an overall rising trend. These signals are effectively removed by the detrending process. Instrumental artifacts like the one highlighted by the orange dashed vertical line can persist after detrending and pose potential challenges in the identification of planetary signals.}
    \label{fig:detrending}
\end{figure*}

\subsubsection{Detrending}

Detecting short timescale and often weak variable signals requires removing systematic trends from the light curves. The T$16$ project applies two detrending methods sequentially: the Trend Filtering Algorithm (TFA, \citealp{TFA_kovacs}) and a Spline-based External Parameter Decorrelation (SEPD). SEPD is applied first, followed by TFA on the SEPD filtered light curves. both procedures are implemented using the \texttt{VARTOOLS} package \citep{VARTOOLS}. 

SEPD detrending is performed with respect to several external parameters: time, CCD position of the source (X and Y) and CCD temperature. Each light curve is modeled as a linear combination of basis functions constructed from these parameters, and the best fit model is subtracted. In practice, this removes low-frequency variations, which are typically dominated by instrumental systematics.

TFA detrending (as described in \citealp{TFA_kovacs}) is a multi-step process. First, a set of template stars is selected: sources that are spatially and magnitude-uniform across the detector, exhibit no strong signs of intrinsic variability in a Lomb-Scargle periodogram, and contain a large number of data points. For each Sector/Camera/CCD combination the T$16$ project constructs a template consisting of $200$ such light curves. A subset of this template is used to derive the filter time base, taken from the light curve with the largest number of  measurements in this subset. Missing data points in individual light curves are filled with the mean value of that light curve. A zero-mean template time series is then computed, with outliers removed via $\sigma$ clipping. The normal matrix of the template is constructed, and each light curve is fitted as a linear combination of the template light curves. The resulting best fit model is subsequently subtracted from each light curve. 

Figure \ref{fig:detrending} illustrates the effect of the detrending pipeline on an example light curve of TIC $273568903$. The top panel shows the raw, undetrended FFI light curve, which displays strong sinusoidal variations and a long-term trend. The bottom panel shows the light curve after passing through the T$16$ detrending pipeline (SEPD+TFA), in which the structures have been effectively removed. Some instrumental artifacts and systematic noise remain, as indicated by the dashed peach vertical line marking a likely instrumental feature that persists after detrending. Such artifacts can mimic transit like signals, particularly in the case of mono-transits (light curves that exhibit only a single transit), which are crucial for identifying long period planets. 

\subsubsection{Resulting Light Curves}

The T$16$ project produced a total of $83,717,159$ detrended light curves for $54,401,549$ stars from TESS Cycle 1 with $T<16$ mag. The number of light curves exceeds the number of targets because the same target may have multiple light curves across various sectors. The TESS-band magnitude T is computed from \textit{Gaia} DR2 photometry following \cite{TESS_mag_Stassun}:
\begin{align}
T &= G - 0.00522555\,(G_{BP}-G_{RP})^3 \notag \\
  &\quad + 0.0891337\,(G_{BP}-G_{RP})^2 \notag \\
  &\quad - 0.633923\,(G_{BP}-G_{RP}) \notag\\
  &\quad + 0.0324473
\end{align}
where stars lacking a \textit{Gaia} color measurement are assigned $T=G+0.0324473$. As noted in \citet{TESS_mag_Stassun}, the precision of this relation decreases for M-dwarfs.

Additional sectors are being processed at the time of writing and will be included in future work. A detailed assessment of the photometric precision of the T$16$ light curves relative to other publicly available pipelines is presented in in \citet{T16_Joel}. In summary, the Signal to Pinknoise achieved by the T$16$ project generally exceeds that of the MIT Quick-Look Pipeline project, the TESS \textit{Gaia} Light Curve project and Goddard Space Flight Center \texttt{eleanor lite} project. While the TESS Science Processing Operations Center Full Frame Image pipeline achieves slightly better precision overall, T$16$ offers the important advantage of including a large number of substantially fainter stars. 

Furthermore, $2639$ out of the $2971$ TESS Objects of Interest (TOIs) known at the time of publication of \cite{T16_Joel} exhibit detectable transit signals in the Cycle $1$ T$16$ light curves, which provides a useful sample to train and test the classification methods applied in this work. 

\section{Classification Method} \label{sec:method}

Machine Learning has transformed exoplanet detection in recent years by enabling the efficient analysis of vast photometric datasets that would otherwise require prohibitive amounts of manual vetting. Numerous methods have been developed and optimized for transit identification including convolutional neural networks (CNNs; e.g. \citealp{Shallue_2018}), random forest classifiers (RFCs; e.g. \citealp{RFC_Classification_Kepler_McCauliff, Pearson_2017_CNN}), and, more recently, transformer-based architectures (e.g. \citealp{Salinas_2025_transformers, Choudhary_2025_Transformer}). Each approach offers specific advantages and drawbacks depending on the structure of the input data, the need for interpretability, and the desired balance between precision and recall when separating planetary signals from astrophysical or instrumental false positives. 

For this work we adopt random forest classifiers as the primary tool for large scale initial classification. RFCs offer a number of practical benefits: They are computationally inexpensive compared to deep learning models, making them well suited for processing millions of TESS light curves; they provide built-in measures of feature importance, facilitating interpretability, and they are robust to noisy, non-linear, and partially missing data. RFCs were introduced by \citet{Breiman_2001_RFCs} and theoretically validated by \citet{RFC_validation_Biau_2010}. They operate by constructing an ensemble of decision trees, each trained on a randomly sampled subset of the data and input features. The aggregate prediction, typically determined by majority vote, reduces overfitting and improves generalization performance. 

In order to train an RFC, each light curve must be represented by a set of diagnostic parameters (the features) that adequately capture the morphology and statistics of transit-like signals. The features must be sufficiently informative for the classifier to learn complex non-linear patterns, but not so numerous that they introduce noise or cause overfitting. To balance these considerations, we employ a two-step classification strategy in which an initial RFC is used to evaluate the relative importance of features, and an optimized subset of these features is then used in the final model. The details of this training procedure are described in Section \ref{sec:training_process}. 

\subsection{Data extraction from detrended Light Curves}

We extract relevant transit parameters from the detrended T$16$ light curves using a multistep procedure. As a first step, the $\sim80,000, 000$ light curves from cycle $1$ are grouped by TESS sector and by their corresponding camera/CCD combination. This organization is essential because blending (flux contamination from nearby sources) typically arises from stars located within the same sector/camera/CCD frame as the target. Targets near detector boundaries may be contaminated by stars lying on adjacent camera/CCD regions. 

After sorting, we apply the Cambridge Exoplanet Transit Recovery Algorithm (CETRA, \citealp{CETRA}) to each light curve. CETRA identifies transits through two steps: a linear search that optimizes transit time and duration, followed by a periodic search that folds the linear-search detection and evaluates the likelihood over a grid of trial periods. CETRA is slightly more accurate than traditional methods such as Box Least Squares (BLS, \citealp{BLS}) or Transit Least Squares (TLS, \citealp{TLS}), but its primary advantage is its GPU implementation, which enables a rapid speed up in transit searches. 

Although CETRA returns accurate period estimates, it does not provide a richer set of statistical parameters required for random forest classification. We therefore perform additional analysis on the CETRA detections using \texttt{VARTOOLS} \citep{VARTOOLS} to compute fixed-period BLS statistics. For each target, we run BLS at the CETRA period as well as at $P/2$, $2P$, and $3P$ to capture harmonic structures that help distinguish planetary transits from eclipsing binaries (EBs). We additionally perform a free period BLS search with a narrow window around the CETRA period to extract further diagnostic features. 

These BLS computations are inexpensive and return $87$ parameters per target. We remove features that are not expected to aid classification, such as raw magnitudes, white noise and red noise estimates and the period itself, and augment the remaining feature set with the stellar effective temperature and luminosity from \textit{Gaia} DR3 when available. This results in a final set of 53 features used to train the random forest classifier, including the effective temperature and luminosity allowing the classifier to learn to reject false positives associated with evolved stars, which are common contaminants in large photometric surveys.

\begin{figure}[ht!]
    \centering
    
    \begin{subfigure}{\linewidth}
        \centering
        \includegraphics[width=\linewidth]{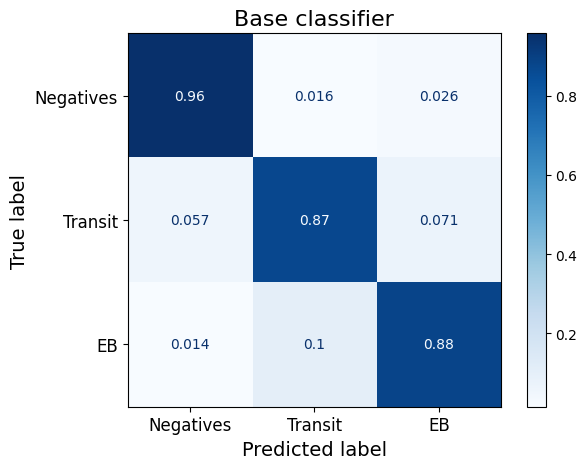}
        \caption{Confusion Matrix for the Random Forest Classifier for all stars. The classifier correctly identifies $87\%$ of the transit systems that are withheld for testing, while $5.7\%$ are labeled as Non-Detections and $7.1\%$ are labeled as EBs.}
    \end{subfigure}
        
    \begin{subfigure}{\linewidth}
        \centering
        \includegraphics[width=\linewidth]{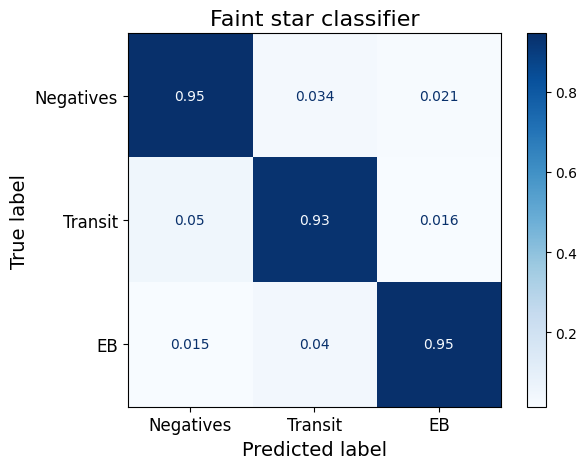}
        \caption{Confusion Matrix for the Random Forest Classifier for faint stars ($T\ge 14.5$). The classifier correctly identifies $93\%$ of transit systems withheld for testing, while $5\%$ are labeled as Non-Detections and $1.6\%$ are labeled as EBs, indicating slightly higher accuracy than the Base classifier.}
    \end{subfigure}
    
    \caption{Confusion Matrices for the Base classifier (all stars, top) and the Faint star classifier ($T\ge14.5$, bottom). Both classifiers are highly effective at identifying transit-like signals from noise, but there are erroneous classifications, many of which are addressed in the subsequent vetting procedure.}
    \label{fig:confusion_matrices}
\end{figure}

\subsection{Random Forest Classifier Training}

Machine learning performance is fundamentally limited by the quality and representativeness of its training data. For transit detection, the training sample must span the range of astrophysical scenarios present in the survey and be sufficiently large for the classifier to learn the relevant patterns. During this project, we found that a single random forest classifier trained on all light curves performed suboptimally across the full magnitude range ($7\lesssim T \lesssim16$). This is primarily because the signal to noise ratio of transit detections strongly depends on the stellar brightness. The noise properties of bright ($T\lesssim 14.5$) and faint ($T\gtrsim 14.5$) stars differ enough that a unified model struggles to learn decision boundaries that generalize across both regimes. 

To address this, we trained two separate classifiers. The first, the base model, was trained on light curves spanning the full magnitude range, but with the training distribution deliberately weighted towards stars with $T<14.5$. The second, the faint star model, was trained exclusively on light curves with $T>14.5$ to find transit candidates around faint stars, which is where we expect to find most of the previously undetected candidates. The division at $T=14.5$ was chosen empirically, classifier performance begins to degrade beyond this threshold when a single unified RFC is used, and the majority of known TOIs are brighter than this limit. 

Both classifiers were trained to distinguish among three classes: ($1$) Negatives (light curves without any significant periodic variability), ($2$) Eclipsing binaries (EBs), and ($3$) planetary transits. We aim to balance the sizes of the three groups in our training sample, and weigh the classes inversely proportional to class-size to further reduce any size related biases.\\

\subsubsection{Training data generation}

Our training data generation procedure is similar for both random forest classifiers, with the primary difference being that TOIs are scarce at faint magnitudes, necessitating a heavier reliance on injected signals for the faint star classifier. For the negative class, we randomly select light curves within the appropriate magnitude ranges. We use $9342$ such light curves for the base model and $5940$ for the faint star model. 

For the EB class we use a combination of real EBs and physically realistic injected signals. We draw the known EBs from the recently published TESS $10,000$ catalog \citep{Kostov_2025_TESS10000}, which includes $7936$ new and $2065$ previously known EB, and from the TESS-EBs catalog \citep{Prsa_2022_TESS_EB}, which includes $4584$ EBs from TESS sectors $1$ through $26$. An EB is only included in our training set if CETRA recovers a period consistent with the catalog value to within $1\%$, ensuring that training examples reflect signals detectable by our pipeline. Using this criterion, we recover $6395$ EBs for our base model classifier and $1160$ for our faint star classifier. 

We augment both EB training samples by injecting EB signals into randomly selected light curves within the respective magnitude bands using \texttt{jktebop} \citep{Southworth_2008_jktebop} via the \texttt{VARTOOLS} program \citep{VARTOOLS}. The stellar parameters of the primaries are retrieved from the TESS Input Catalog (TIC) \citep{Stassun_2018_TIC}, and companion properties are sampled using the MIST stellar evolution models \citep{Paxton_2011_MIST_1, Paxton_2012_MIST_2, Paxton_2015_MIST_3, Dotter_2016_MIST_4, Choi_2016_MIST_5}. The mass ratio is sampled uniformly between $0.2$ and $1.5$, the orbital separation log-uniformly between $5-23 R_\odot$ and the inclination uniformly between $85\degree$ and $90\degree$. Orbital periods are computed from Kepler's third law. As before, injected systems are only retained if CETRA recovers the period to within $1\%$ accuracy. We thus add an additional $2309$ injected EB signals for the base model training sample and $7298$ for the faint star model.

A similar strategy is used to construct the planetary transit class. We first identify TOIs whose transit signals are successfully recovered by CETRA, again requiring agreement within $1\%$ of the catalog period. For the base model we find $3632$ such light curves from $2095$ individual TOIs, but for the faint star model we only recover $91$, reflecting the scarcity of faint TOIs. To compensate, we inject transit signals into randomly selected light curves. Planetary parameters (radius, mass, and orbital period) are sampled from the TOI catalog, while orbital inclination and phase are drawn randomly. The eccentricity is fixed to $0$ and a quadratic Limb-Darkening law with parameters $u_1=0.3471$ and $u_2=0.3180$ is adopted. Transits are injected using the \texttt{injecttransit} command in \texttt{VARTOOLS} \citep{VARTOOLS}. As with EBs, only injections recovered by CETRA to within $1\%$ of the input period are included. Using this procedure we add $1456$ planetary signals for the base model and $8517$ for the faint star model. 

Because the relative abundances of Negatives, EBs, and Transits differ in both training sets, both classifiers are trained using class-weighted loss functions, with weights inversely proportional to class frequency. This ensures that minority classes contribute proportionally to model optimization and prevents the classifier from being biased toward the more common training examples.

\subsubsection{Training process} \label{sec:training_process}

We train our random forest classifiers twice for each model (faint and base). Throughout this project we use Python's \texttt{Scikit-learn}'s random forest implementation \citep{scikit-learn}. During the first training step we feed in all the parameters that we expect could benefit the classification and perform a grid search for hyperparameter tuning. We extract the top $30$ features and then retrain the classifier on only those features. We found that doing this helps prevent overfitting while still resulting in good scores. We randomly selected $20\%$ of the training data to withhold from the modeling for testing purposes. Our model achieves a weighted average f1-score of $0.91$, with receiver operator curve (ROC) areas under the curve (AUC) of $0.99$ for the classification of non-detections and $0.98$ for both transit and EB classifications. The confusion matrix for our RFC is shown in Figure \ref{fig:confusion_matrices} for both the faint and base model classifier. Both classifiers found the signal-to-pinknoise of the primary CETRA peak to be the most important classification feature, followed by the signal-to-pinknoise at multiples of the period returned by CETRA and the signal residues at these same periods. 

An RFC assigns each light curve a probability of belonging to each class, meaning that a "transit detection" could in principle occur at probabilities as low as one third (equivalent to random guessing among the three classes). To choose an appropriate probability threshold for retaining transit candidates, we manually vetted around $1500$ light curves spanning multiple sectors for both models and compared the human classifications against the RFC outputs as a function of probability cutoff for keeping a transit detection. We adopted a minimum transit probability of $0.5$, which provides a favorable balance between false positives and negatives. This threshold is also intuitive, for transit signals above this threshold are more likely to be transits than not. 

\subsection{Post RFC automated vetting}\label{Sec:automated_vetting}

After classification, we combined the outputs of the two RFCs and process each sector/ccd combination individually. The automated vetting begins by removing candidates affected by local contamination, which is a significant issue in TESS data due to the large $20''$ pixel scale. Using the \texttt{NetworkX} package \citep{networkx} we construct a graph in which nodes correspond to all sources classified as either transit-like or EB-like, and edges connect pairs of sources that ($1$) lie within two arcminutes of each other, and ($2$) share compatible ephemerides and periods. Each connected component represents a contamination group. Within each group, we retain only the source exhibiting the strongest transit signal and discard the remainder as likely blends. 

We next impose a signal-to-pinknoise \citep{Pont_2006_Pinknoise} cutoff of $S/N_{pink} > 10$, which we find provides best balance between rejecting spurious detections and retaining genuine transit candidates. After this cut, we examine the period distribution of the surviving sources. Instrumental or detrending related systematics often recur at characteristic periods and ephemerides across a given CCD, therefore, when a period bin is overly populated and the associated candidates also share similar transit times, we remove those signals as likely systematics. 

For all remaining candidates, we fit a planetary transit model to the remaining light curves using the \texttt{BATMAN} package \citep{BATMAN_Kreidberg_2015}. Any candidate whose best fit planetary radius exceeds $2.5R_J$ is discarded as a likely EB. This threshold is intentionally generous, additional large radius false positives are filtered later in the manual vetting step. The \texttt{BATMAN} fit performed here is a fast, initial fit that is applied to all candidates that pass our filters up to this stage. This initial fit often fails, leading to underestimations of planetary radii in the vetting plots. This is accounted for in the manual vetting process. 

\begin{figure}
    \centering
    \includegraphics[width=\linewidth]{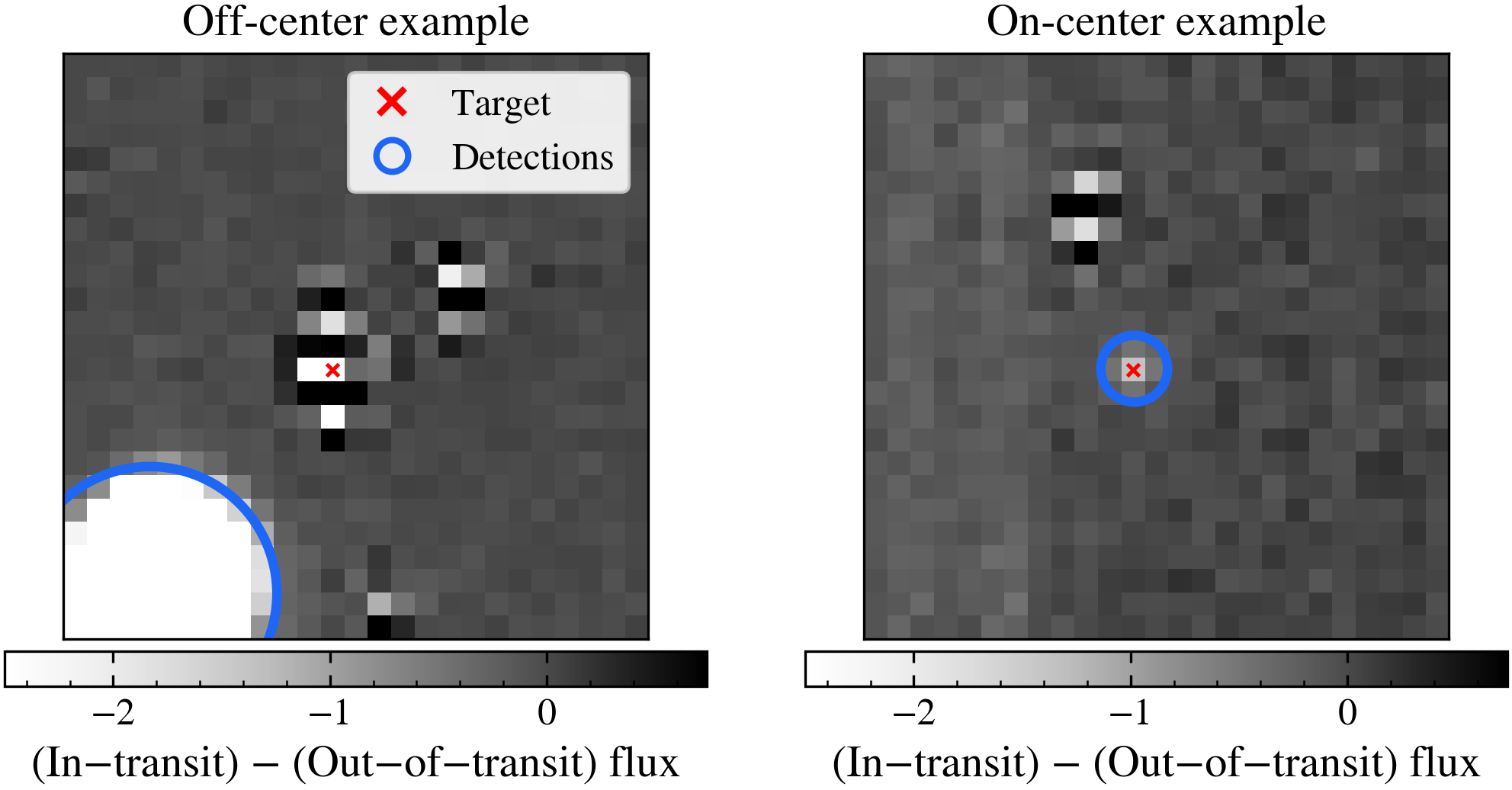}
    \caption{\textit{Left}: Example of automated off center source detection. This example shows an off center residual indicating a blend as the source of the measured variability. The checkered pattern at the source position is consistent with Poisson noise or imperfect image subtraction due to the brightness of the source. \textit{Right}: Localized, on center residual indicating that the variability is due to the target.}
    \label{fig:StarFinder}
\end{figure}

We also incorporate information from the background subtracted FFIs into our vetting process. For each candidate, we extract a $25\times25$ pixel cutout centered on the target and compute the difference between stacked in-transit and out-of-transit subtracted images. A genuine transit should manifest as a localized decrease in flux at the target position, with minimal residuals everywhere else. Using Astropy's \texttt{DAOStarFinder} package \citep{DAOStarFinder} we run a source detection on this in-transit minus out-of-transit cutout image (after inverting to ensure a positive source) and remove candidates for which the transit source was detected to be off-center, as these are most likely blended. These cutout images also play an important role in our manual vetting process, as will be described later. Figure \ref{fig:StarFinder} shows examples of our source detection: In the off-center example on the left a strong signal appears away from the target location, seemingly originating from a nearby star. The target position displays a checkered pattern, consistent with imperfect subtraction of relatively bright sources. In contrast, the on-center example displays a well-localized decrease in flux consistent with a real transit. If the StarFinder does not detect any sources we do not filter the corresponding source out since the transit depth in this case may simply be too shallow for automated detection. 

The last step of our automated vetting process consists of a second round of RFC vetting designed to mimic manual vetting. To train this classifier, we manually vetted and classified approximately $4,500$ candidates. Signals identified as blended systems were excluded from the training set, as blends are flagged using direct image information that is not available to the classifier. The RFC performs a binary classification between transit-like and non-transit-like signals using the same set of features as the classifiers described above. We trained the model on $80\%$ of the labeled sample and reserved the remaining $20\%$ for validation. The classifier was tuned to maximize recall in order to minimize the loss of true transit signals. We adopt a conservative transit probability threshold of $0.2$ for a signal to be classified as a transit. With this choice, the classifier removes approximately half of the signals classified as false positives in the validation set while retaining $98\%$ of the transit signals identified through manual vetting.

Before applying this second stage classifier we separate mono-transit candidates (light curves with only one transit like dip) from the other candidates. We do not apply this RFC to monos because one of its most informative features, ``fraconenight", which describes the fraction of total $\Delta\chi^2$ signal contributed by a single transit, does not apply meaningfully when only a single transit is present. Since monos constitute only a small fraction of the total candidate pool, we opt to vet them manually rather than develop a dedicated classifier.

\begin{figure*}
    \centering
\includegraphics[width=\linewidth]{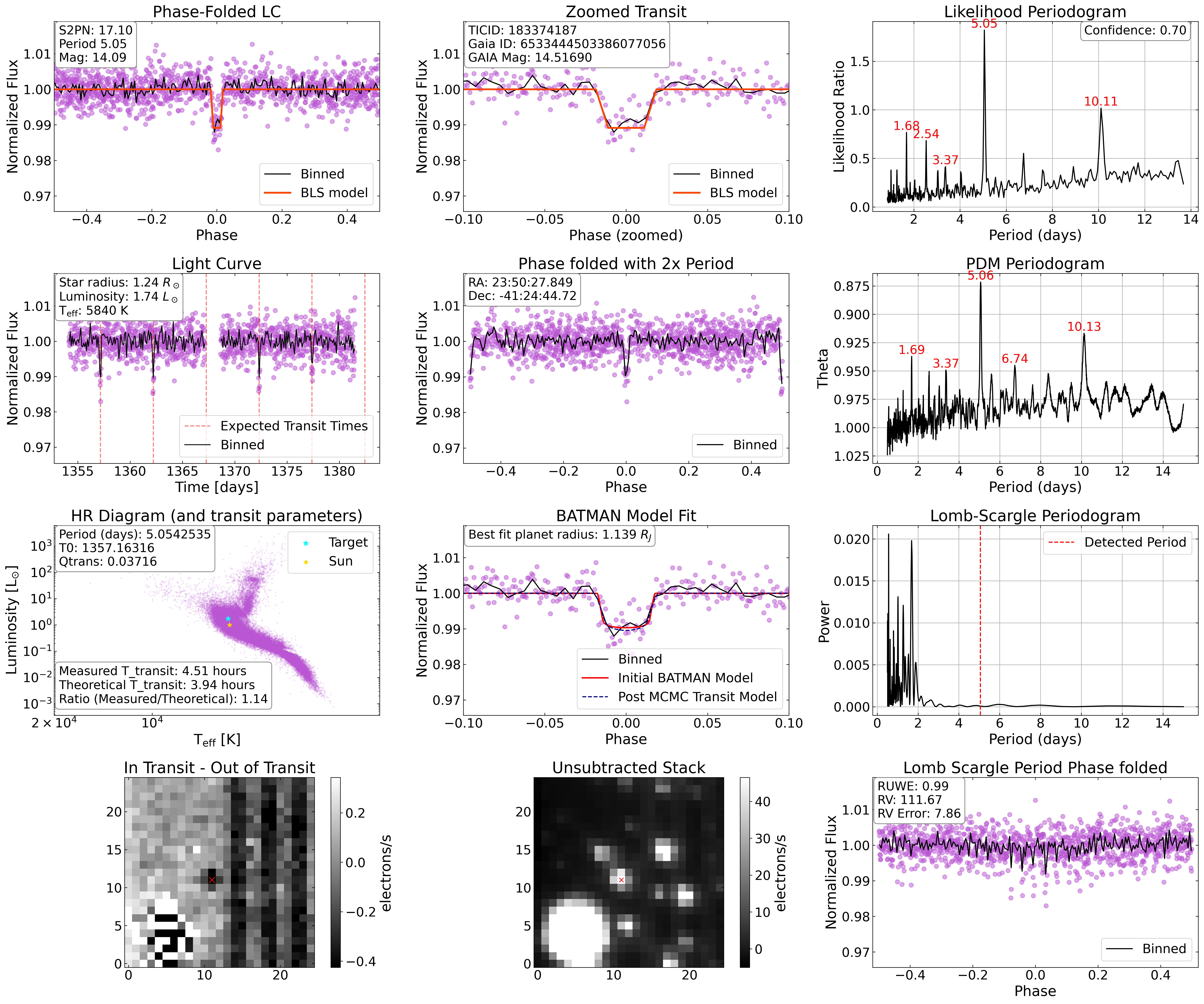}
    \caption{Example vetting ``summary'' plot for the now confirmed exoplanet TIC $183374187$. Detailed description of each panel is provided in the main text. The complete figure set contains one image for each of the $11,143$ objects in Table \ref{tab:parameter_descripton} and the $411$ single transit candidates, and is accessible both through the online journal and via https://doi.org/10.7910/DVN/CWEUGW.}
    \label{fig:vettingPlot}
\end{figure*}

\subsection{Manual Vetting}

Although machine learning algorithms and automated vetting procedures are highly effective for processing large datasets, they do not yet match human performance in edge cases or in low signal to noise regimes. Therefore, the final step of our transit search consists of manual vetting, and every candidate we publish has been inspected by at least one vetter. For this purpose, we generate standardized vetting plots, an example of which is shown in Figure \ref{fig:vettingPlot}. The panels included in these plots are described below.

The top left panel displays the detrended lightcurve phase folded using the period returned by CETRA. Violet points represent the unbinned data at the $30$ minute cadence of TESS cycle $1$ FFIs, the black curve shows the data binned in 200 phase bins, and the red curve shows the best fit BLS model. The \texttt{VARTOOLS} BLS implementation fits a trapezoidal model rather than a simple box, allowing for finite ingress and egress durations and thereby improving the fit. 

The top middle plot shows a zoomed in view of the transit region to highlight the transit morphology.
The top right panel shows the CETRA power spectrum returned with the five highest peaks labeled.

The left panel in the second row shows the full detrended light curve (unbinned in violet and binned in black), with red dotted lines marking the expected transit times returned by CETRA. The middle panel phase folds the light curve at twice the CETRA period. A depth asymmetry between the events at phase $0.0$ and $\pm0.5$ would hint at that candidate being an EB rather than a planetary transit. 

The third panel in this row presents the phase dispersion minimization (PDM) of the candidate \citep{PDM_Stellingwerf_1978}. The PDM method consists of folding the light curve over several trial periods and binning it into $M$ phase bins. The phase-folded dispersion is then calculated and compared to the overall dispersion. This is quantified via the $\theta$ parameter plotted on the y-axis of the panel, which can be evaluated using the following equation:
\begin{equation}
    \theta(P) = \frac{\sum_{j=1}^{M}{(N_j-1){s_j}^2}}{(N-1)s^2}
\end{equation}
where $P$ is the period, $s_j$ is the variance of bin $j$ containing $N_j$ points, and s is the total variance of the dataset. Thus, $\theta(P)$ measures the ratio of phase-binned scatter to overall scatter, ranging from $0-1$. Values near one indicate no periodicity while a lower $\theta$ marks trial periods for which the data aligns well, indicating periodicity. Unlike Fourier-based methods, PDM assumes no specific signal shape and is therefore well suited for eclipsing, pulsating, and other non-sinusoidal signals. We compute the PDM using the \texttt{pyPDM} implementation from the \texttt{PyAstronomy} toolkit \citep{pyastronomy} using a period grid consisting of equally spaced periods from $0.5$ to $15$ days at a resolution of $0.005$ days. For each trial period we divide the light curve into 50 equally spaced phase bins. To reduce sensitivity to binning artifacts, we adopt 5 different phase offsets and average the results. The top five minima are marked in the figure.

The left panel of the third row shows the position of the source on a Hertzsprung-Russell diagram, with the Sun shown for reference. Stellar parameters are taken primarily from the TESS Input Catalog (TIC, \citealp{Stassun_2018_TIC}), supplemented by \textit{Gaia} DR3 \citep{Gaia_DR3} when necessary. This diagnostic is used to rapidly identify giant stars and to assess whether the measured transit depth is physically plausible. The panel also lists the CETRA period, transit epoch, and the measured transit duration as a fraction of the orbital period.

We also compute the expected maximum transit duration from Kepler's law, 
\begin{equation}
        T_{transit} = \left( {\frac{3P}{\pi^2\rho G}}\right)^{\frac{1}{3}} 
\end{equation}
and compare this value to the BLS derived transit duration. Transit durations significantly longer than expected typically indicate EBs or blended systems, while unusually short durations may arise from grazing geometries. However, we do not impose a strict cutoff on the ratio of measured to predicted duration mainly due to uncertainties in the stellar density and also the possibility of eccentric orbits.

The middle panel in the third row mirrors the zoomed in phase folded view shown earlier but with the best fit \texttt{BATMAN} transit model \citep{BATMAN_Kreidberg_2015} overplotted in red. The inferred planetary radius (in Jupiter radii) is listed in the corner, aiding in distinguishing planetary candidates from EBs. In general, we remove candidates with inferred planetary radii greater than $2.4R_J$. As mentioned in Section \ref{Sec:automated_vetting}, the \texttt{BATMAN} fits are often incorrect. Part of the manual inspection is to assess cases where the modeling failed and to estimate the planetary radius manually. The dashed navy line represents the post MCMC transit fits described in Section \ref{Error_estimation}, which we added post-vetting to make clear that the initial \texttt{BATMAN} fit was not used to estimate transit parameters.

The last panel in the third row shows the Lomb-Scargle (LS) periodogram, which is used to identify sinusoidal variability, which can indicate stellar activity or other non-transit related signals. Although transit signals may also appear in the LS periodogram, strong sinusoidal signals in combination with other indicators can aid in identifying false positives.

The first panel in the fourth row shows a $25\times 25$ pixel cutout of the stacked, subtracted image in transit minus out of transit. A genuine transit should produce a localized, on-center, negative residual. 

The middle panel shows the corresponding unsubtracted stacked image, which provides context for the stellar environment. These two panels together allow the vetter to confirm whether the signal originates from the target or a blended neighbor. In the example of Figure \ref{fig:vettingPlot}, the target (red cross) is centered on a source that shows clear in transit dimming. The residual vertical stripes in the left panel are a regularly repeating set of three stripes on the front side of the TESS CCDs designed to improve readout speed \citep{TESS_Instrument_Handbook}. Bright neighboring sources like the one in the lower left corner of the images often leave a spotted pattern due to Poisson noise, which does not indicate variability. 

The final panel shows the light curve phase folded at the period corresponding to the highest peak in the LS periodogram. Significant periodic variability at this period indicates the presence of a coherent signal, and the phase folded light curve allows us to assess the shape and modulation of this signal. A sinusoidal or quasi-sinusoidal shape could suggest stellar variability like rotation or pulsations. In addition, we list certain \textit{Gaia} DR3 astrometric and radial velocity parameters when available. These include the \textit{Gaia} renormalized unit weight error (RUWE), median RV, and RV uncertainty. RUWE serves as a quality indicator of the astrometric solution and can hint at unresolved hierarchical triples as astrophysical false positives. \textit{Gaia} RVs, though limited to relatively bright FGK stars, provide additional evidence for or against binarity and help identify likely false positives. 

\subsection{Error Estimation}\label{Error_estimation}
While the algorithms we employ in our transit search, namely CETRA and BLS, are effective at detecting transit signals, they do not provide uncertainties on the inferred transit parameters. It is, however, imperative to quantify credible uncertainties, particularly on the ephemeris and orbital period of transiting planet candidates, since accurate prediction of future transit windows is essential for scheduling follow-up observations. 

To estimate parameter uncertainties, we fit a Mandel–Agol transit model \citep{Mandel_Agol_2002_transitModel} to each candidate light curve using the \texttt{BATMAN} package \citep{BATMAN_Kreidberg_2015}. As opposed to the \texttt{BATMAN} fits described in Section \ref{Sec:automated_vetting}, this round of fitting is much slower to get accurate parameters and uncertainties. We sample the posterior with an affine-invariant ensemble Markov Chain Monte Carlo (MCMC) sampler implemented in the \texttt{emcee} package \citep{Foreman-Mackey_2013_emcee}, which follows the algorithm of \citet{Goodman_2010_MCMC}.

We apply a single $7\sigma$ clipping pass to remove strong outliers and convert the detrended TFA magnitudes to normalized relative flux. Because the formal per-point uncertainties provided in the T16 light curves may be underestimated, we adopt a white-noise model with a single scalar dispersion $\sigma$, defined as the standard deviation of the out-of-transit (OOT) flux. 

We fit for the orbital period, the mid-transit time (of the first transit), the planet-to-star radius ratio, the scaled semi-major axis ($a/R_\star$), the orbital inclination, and the two limb-darkening coefficients. We use the quadratic limb-darkening law first introduced by \citet{Kopal_1950_quadr_limb_dark}:
\begin{equation}
    I(\mu)/I(1) = 1 - u_1(1 - \mu) - u_2(1 - \mu)^2,
\end{equation}
where $\mu = \cos(\theta)$ and $\theta$ is the angle between the line of sight and the local stellar surface normal. We adopt the parameterization of \citet{Kipping_2013_LimbDarkening}, which redefines the limb-darkening coefficients as
\begin{equation}
    q_1 = (u_1 + u_2)^2,
\end{equation}
and
\begin{equation}
    q_2 = \frac{u_1}{2(u_1 + u_2)},
\end{equation}
allowing uniform priors while ensuring physically meaningful combinations of $u_1$ and $u_2$. We assume circular orbits ($e = 0$) for all candidates. The \texttt{BATMAN} light curves are computed using a supersampling factor of 3 and an exposure time of 30 minutes, matching the cadence of the TESS full-frame images (FFIs) in Cycle 1.

Given the OOT-derived scatter $\sigma$, we use a Gaussian likelihood with constant variance:
\begin{equation}
    \ln \mathcal{L}(\boldsymbol{\theta}) = -\frac{1}{2}\sum_{i=1}^{N}
    \left[\frac{f_i - m(t_i|\boldsymbol{\theta})}{\sigma}\right]^2
    - \frac{N}{2}\ln(\sigma^{-2}),
\end{equation}
where $N$ is the number of data points and $m(t_i|\boldsymbol{\theta})$ is the model flux. We adopt simple uniform (box) priors on the varied parameters, centered around the values obtained from the BLS analysis.

We initialize 32 walkers in a small Gaussian ball around the starting parameter vector and discard an appropriate burn-in phase. Every 1000 steps, we compute the integrated autocorrelation time $\tau$ for each parameter to track convergence, and continue sampling until the slowest parameter has at least $\sim 50\tau$ effective samples or a cap of 25,000 steps is reached. This procedure yields well-mixed posterior distributions and ensures robust convergence of the ephemeris and period parameters for most candidates.

For each target, we flatten the chains and report the posterior median along with the 16th and 84th percentiles as the $1\sigma$ credible intervals. These uncertainties are included in our final candidate summary table (Table \ref{tab:parameter_descripton}), and are also reported to ExoFOP as part of our CTOI submissions. The best fit transit models are furthermore added to our vetting plots in post processing (see Figure \ref{fig:vettingPlot}).
\section{Results} \label{sec:results}
\subsection{Candidate Catalog}
Our automated pipeline reduced the initially more than $80,000,000$ light curves to $\sim 50,000$ transiting exoplanet candidates, which we manually vetted. After cross-referencing all signals that passed our automated and manual vetting procedure with the TESS EB catalog \citep{Prsa_2022_TESS_EB}, the TESS $10000$ catalog \citep{Kostov_2025_TESS10000}, and VizieR \citep{10.26093/cds/vizier,vizier2000} $13,880$ signals remain that are not listed as known EBs in these catalogs. Out of these signals $11,554$ are from distinct sources. Thus, our resulting candidate catalog consists of $11,554$ planet candidates, out of which $411$ are single transits, for which we only note the transit times, depths, and durations returned by our BLS fit. We do not attempt to estimate parameter uncertainties for these candidates (see Table \ref{tab:singles_parameter_descripton}). For candidates appearing in multiple sectors we choose the sector with the highest signal to pinknoise ratio. Alternatively, one could treat repeated candidates jointly, which could aid the detection of longer period candidates and may root out some astronomical false positives. We do not take this approach in this project, but future searches will include this feature. The catalog with all our candidates will be available in machine readable format as supplementary material to this paper and all candidates will be published as community TESS objects of interest (CTOIs, \citealp{Exoplanet_archive_Christensen}). The parameters along with short descriptions are listed in Table ~\ref{tab:parameter_descripton}. For our catalog of monos we solely include the \textit{Gaia} DR3 IDs, the TESS IDs, sectors, cameras and the CCDs in which the monos were detected since we are unable to meaningfully constrain the transit parameters. 

\begin{figure}[h]
    \centering
    \includegraphics[width=\linewidth]{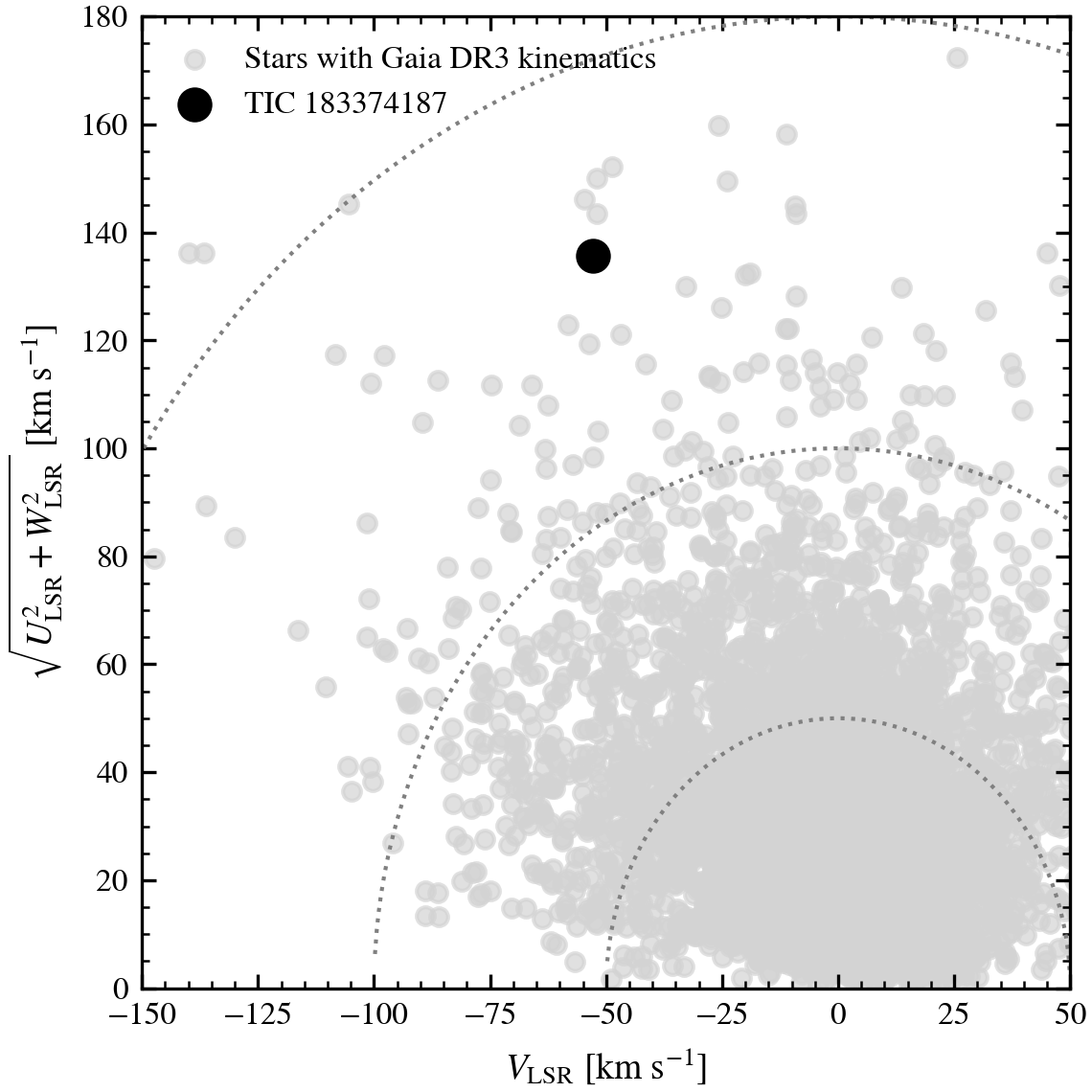}
    \caption{Toomre diagram showing the Galactic space velocities of TIC $183374187$ relative to the local standard of rest (LSR). The star's velocity components ($U_{LSR}$,$V_{LSR}$,$W_{LSR}$) were computed using \textit{Gaia} DR3 astrometry and radial velocity. The dashed curves indicate loci of constant total space velocity. Grey points indicate kinematics of the other targets in my sample with a \textit{Gaia} DR3 radial velocity measurements. The position of TIC $183374187$ indicates kinematics consistent with membership in the Galactic thick disk. This interpretation is supported further by spectral chemical abundance measurements.}
    \label{fig:Toomre}
\end{figure}

\begin{deluxetable}{cccccc}
\tablecaption{Radial Velocity Measurements\label{tab:rv}}
\tablehead{
\colhead{BJD$_{\rm TDB}$} &
\colhead{RV} &
\colhead{$\sigma_{\rm RV}$} \\
\colhead{(days)} &
\colhead{(m s$^{-1}$)} &
\colhead{(m s$^{-1}$)} &
}
\startdata
2460861.82858 &   14.32 & 12.67 \\
2460865.80050 &   74.04 & 10.24 \\
2460873.81564 &    0.00 & 11.44 \\
2460923.79236 &  -68.36 & 12.26 \\
2460923.81117 &  -20.47 & 12.61 \\
2460928.75685 &  -28.98 & 11.91 \\
2460929.76255 &   72.41 & 11.50 \\
\enddata
\end{deluxetable}

Out of our $11,554$ TEP candidates, $928$ are listed as TESS objects of interests (TOIs) and $124$ as community TESS objects of interest (CTOIs) at the time of writing (January $2026$). It is worth noting that the TOI and CTOI catalogs currently consist of $7,890$ and $3,926$ candidates respectively, but only $3, 549$ of these TOIs orbiting $ 3400$ unique targets are listed with cycle $1$ light curves. The number of TOIs recovered must not be used to make arguments related to the completeness of our sample, but the fact that many of our candidates are previously undetected is promising and suggests that this work will increase the sample of transiting TESS objects substantially. Our pipeline's aptitude at recovering TOIs is further discussed in Section \ref{sec:TOI}

\section{Confirmation of TIC $183374187$}\label{sec:confirmation}

\begin{figure}[h!]
    \centering
    \includegraphics[width=\linewidth]{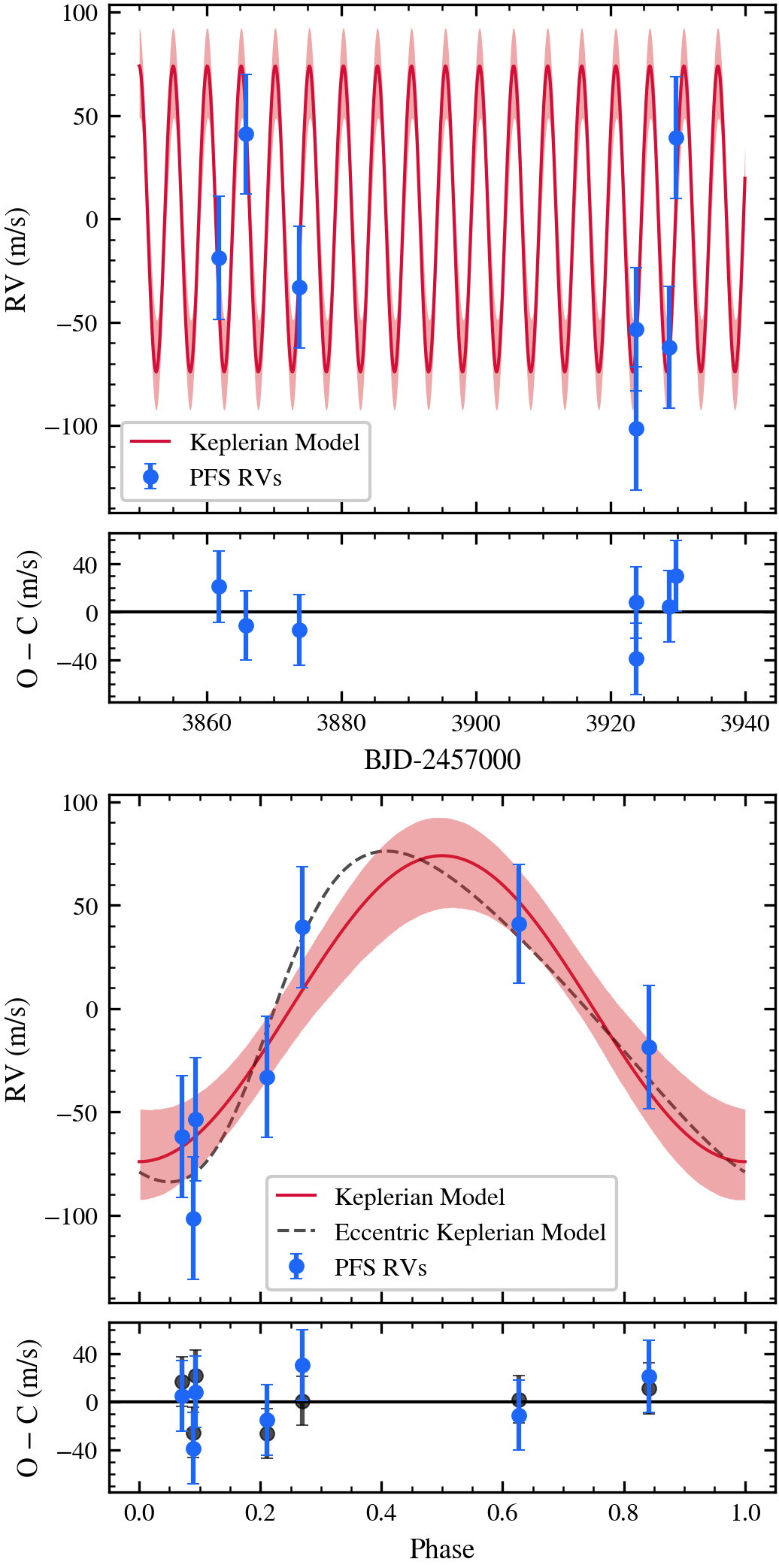}
    \caption{The top panel shows the $7$ radial velocity data points along with the best fit Keplerian model and the corresponding $1\sigma$ uncertainty region in red. Plotted directly below are the residuals. Similarly, the bottom panel shows the phase folded equivalent, also with corresponding residuals. The best fit eccentric Keplerian model is shown as dashed line, its residuals are plotted as black data points.}
    \label{fig:RV_vels}
\end{figure}

As a proof of concept we followed up one of our candidates, TIC $183374187$ (\textit{Gaia} DR3 $6533444503386077056$), a metal poor, solar type star for which \textit{Gaia} DR3 reports a radius of $1.29~R_\odot$ and luminosity of $1.74~L_\odot$, which we kinematically determined to be a member of the thick disk. We calculated the UVW velocities using \texttt{PyAstronomy} \citep{pyastronomy}, the resulting Toomre diagram is shown in Figure \ref{fig:Toomre}. As kinematics alone are not sufficient to confirm population membership \citep{Ferreira_2025_alpha_elements}, we took advantage of the high-resolution mode ($R = 110,000$) of the Planet Finder Spectrograph (PFS) to measure the abundances of the $\alpha$ elements \ion{Mg}{1}, \ion{Si}{1}, \ion{Ca}{1}, and \ion{Ti}{1} and \ion{Ti}{2}. The abundances were derived using the Qoyllur-quipu ($\mathrm{q}^{2}$) Python code \citep{Ramirez:2014A&A...572A..48R}, which relies on the Kurucz's \textsc{ODFNEW} model atmospheres \citep{Castelli:2003IAUS..210P.A20C} and the 2019 version of the local thermodynamic equilibrium (LTE) radiative transfer code \textsc{MOOG} \citep{Sneden:1973PhDT.......180S}. Equivalent widths were measured manually for a selected set of lines, with deblending performed when necessary using Gaussian fits to the line profiles. The full methodology is described in \citet{Ferreira_2025_alpha_elements} and \citet{Galarza_2025_alpha_elements}. We obtained [Mg/Fe] = $0.297 \pm 0.135$ dex and [$\alpha$/Fe] = $0.302 \pm 0.106$ dex, where the latter represents the average of the four $\alpha$ elements. The quoted uncertainties include both observational errors and those arising from uncertainties in the stellar parameters. The high values of [Mg/Fe] and [$\alpha$/Fe] strongly support thick-disk membership for TIC 183374187.

The star is moderately faint, its \textit{Gaia} G band magnitude is $14.5169$. Using the PFS, a high resolution echelle spectrograph on the $6.5$ meter Magellan Clay telescope at Las Campanas Observatory, Chile \citep{Crane_2006_PFS, Crane_2008_PFS, PFS_Magellan}, we obtained seven RV data points over two observing campaigns, shown in Table \ref{tab:rv}. The instrument contains an iodine absorption cell, which provides precise wavelength calibration and an estimate of the instrument point spread function \citep{Geoffrey_Butler_1992,Butler_1996}. 

\begin{deluxetable}{ccc}
\tablecaption{TESS Light Curve Measurements\label{tab:lc}}
\tablehead{
\colhead{BJD$_{\rm TDB}$} &
\colhead{TFA} &
\colhead{$\sigma_{\rm TFA}$} \\
\colhead{(days$-2457000$)} &
\colhead{(mag)} &
\colhead{(mag)}
}
\startdata
1354.14065 & 14.09018 & 0.00356 \\
1354.16148 & 14.08879 & 0.00356 \\
1354.18231 & 14.08985 & 0.00356 \\
1354.20315 & 14.09142 & 0.00356 \\
1354.22398 & 14.08987 & 0.00356 \\
1354.24482 & 14.08762 & 0.00356 \\
1354.26565 & 14.09404 & 0.00356 \\
1354.28648 & 14.08468 & 0.00356 \\
1354.30732 & 14.08219 & 0.00356 \\
1354.32815 & 14.09500 & 0.00356 \\
\enddata
\tablecomments{
Only the first 10 rows are shown.
The full light curve is available in machine-readable form in the electronic edition of the journal.
}
\end{deluxetable}

We perform a joint analysis of the PFS RV data, our detrended T$16$ light curve for TIC $183374187$ (the first ten rows of which are shown in Table \ref{tab:lc}), \textit{Gaia} DR3 astrometric data, and available broad-band photometry. We followed the procedure described in \cite{Hartman_2019_lfit}, which uses a Keplerian orbital model for the RV data coupled with a quadratic limb darkened Mandel Agol transit model for photometric data \citep{Mandel_Agol_2002_transitModel}. We furthermore constrain the stellar parameters using broad-band photometry from \textit{Gaia}, 2MASS \citep{Skrutskie_2006_2MASS}, and WISE \citep{Wright_2010_WISE}, \textit{Gaia} DR3 parallax and stellar atmospheric parameters which were derived spectroscopically to be $T_{eff}=5770\pm60$\,K, $\log\text{g}=3.92\pm0.08$\,dex, $\text{v}\sin{i}<2\,\mathrm{km\,s^{-1}}$, and [Fe/H]$=-0.40\pm0.06$\,dex using the \texttt{SpecMatch-Synth} procedure \citep{Petigura_2015_specMatch_synth, Petigura_2017_specMatch}. During the fitting procedure, the macroturbulent broadening was fixed according to Equation 1 in \citet{Valenti_Fischer_2005}. Microturbulent broadening, together with other intrinsic sources of line broadening such as thermal and collisional effects, is incorporated into the model atmospheres and line formation used to generate the synthetic spectra in the \citet{Coelho_2005} library, and was therefore not treated as an independent free parameter in the fitting.

Throughout the joint modeling we constrain the stellar parameters to be consistent with MIST v1.2 isochrones \citep{Choi_2016_MIST_5,Dotter_2016_MIST_4}, computed with the MESA stellar evolution code \citep{Paxton_2011_MIST_1,Paxton_2012_MIST_2,Paxton_2015_MIST_3}. Note that these models are computed assuming a solar abundance pattern, however we do find evidence for $\alpha$-element enhancement in the spectrum. This discrepancy may lead to modest errors in the inferred mass, radius and age of the host star, and consequently in the mass and radius of the planet. We placed a Gaussian prior on the line-of-sight extinction $A_V$ and set a maximum allowed extinction using the MWDUST 3D Galactic extinction model \citep{Bovy_2016_MWDUST}. Additionally, we placed Gaussian priors on the limb-darkening coefficients used in the \cite{Mandel_Agol_2002_transitModel} models based on theoretical calculations by \cite{Claret_2012_LimbDarkening, Claret_2013_LimbDarkening} and \cite{Claret_2018_LimbDarkening}.

\begin{figure}
    \centering
    \includegraphics[width=\linewidth]{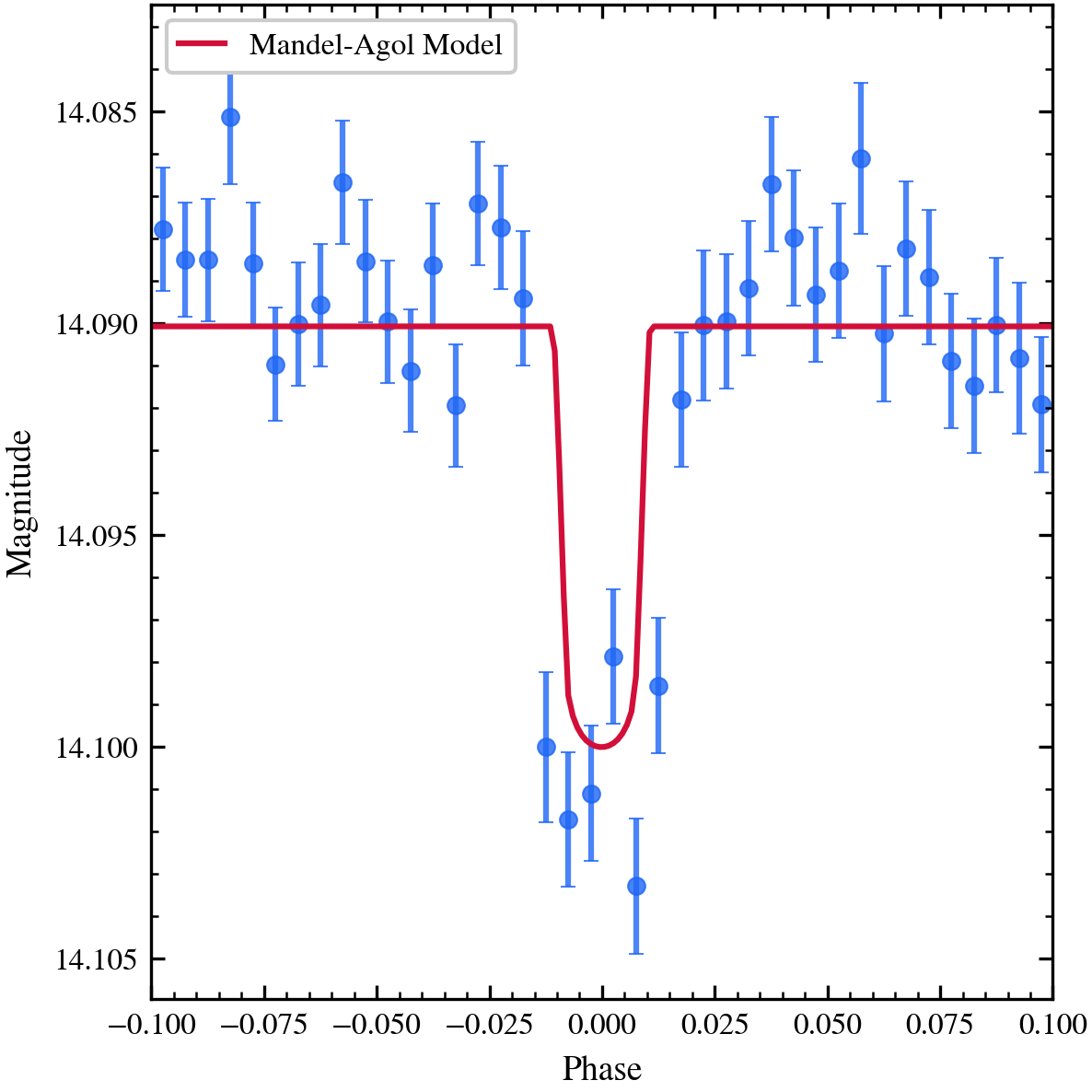}
    \caption{The phase folded T$16$ light curve, measured in TESS sector 2, camera 2 ccd 4 in magnitude space. The data has been binned into 200 phase bins. Overplotted in red is the best fit Mandel-Agol Model, fitted using the BATMAN package \citep{BATMAN_Kreidberg_2015}.}
    \label{fig:transit}
\end{figure}

We estimated the planetary, orbital, and stellar parameters using a Differential Evolution Markov Chain Monte Carlo (DEMC) algorithm \citep{Braak_2006_DEMC}, which is advantageous compared to conventional MCMC due to its speed of calculation and conversion. Before starting the DEMC chain we run a downhill simplex \citep{NelderMead_1965} to obtain an initial best-fit solution, which is then refined via the DEMC sampler. We ran the DEMC sampler until convergence, ensured by visual inspection of the chain traces, of which we discarded an adequate number as burn in. 

\begin{figure}
    \centering
    \includegraphics[width=\linewidth]{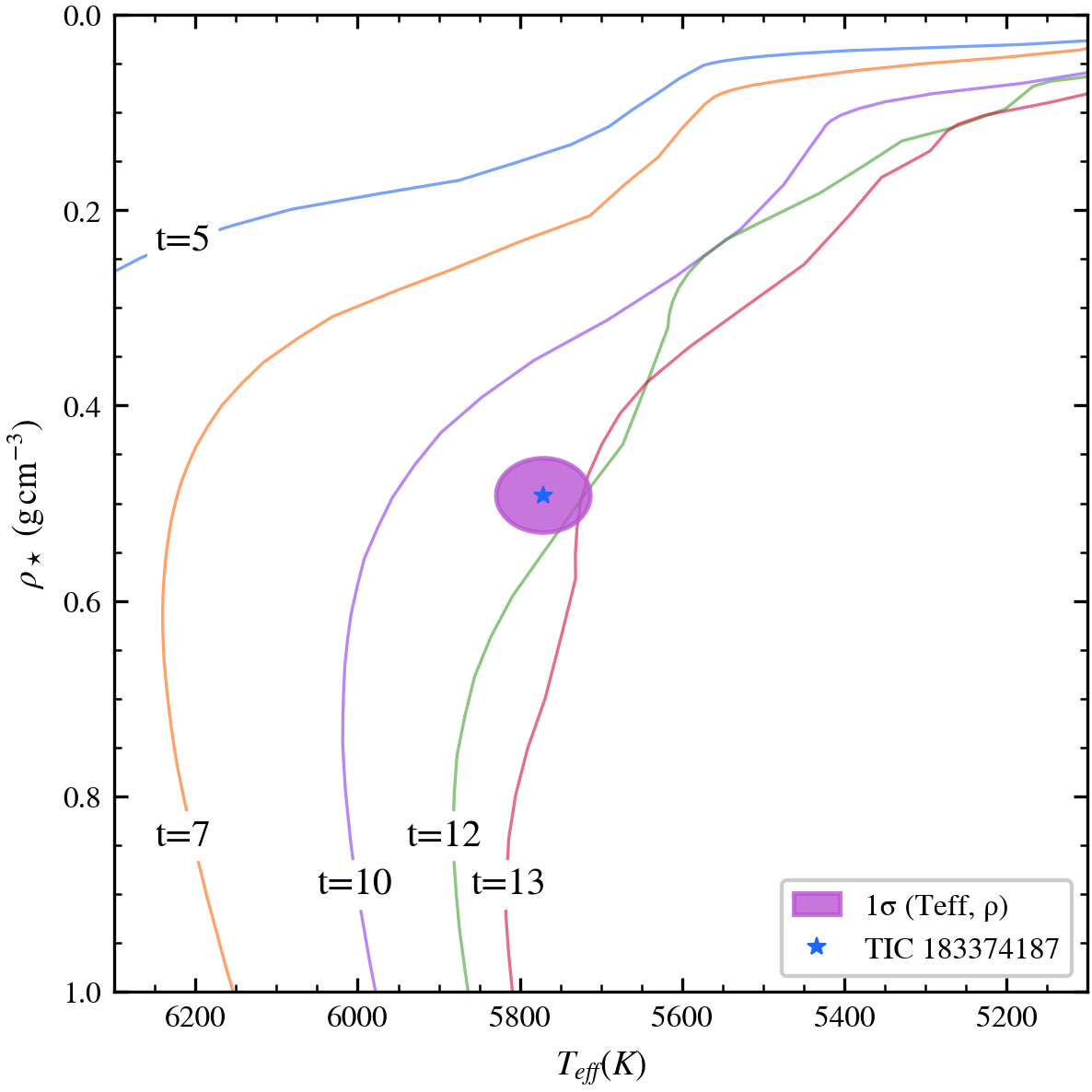}
    \caption{Position of TIC 183374187 on the stellar density versus effective temperature plot, along with our $1\sigma$ uncertainty in violet. Plotted in various colors are stellar isochrones for the inferred metallicity of TIC $183374187$, which were obtained via MIST. The ages are in Gyrs. It is apparent that TIC 183374187 is near the turnoff of the $12$ and $13$Gyr isochrones, hinting at an old, slightly evolved host star.}
    \label{fig:isochrones}
\end{figure}

We fit both a circular orbit model and an eccentric model, where we vary the parameters $\sqrt{e}\cos\omega$ and $\sqrt{e}\sin\omega$. e is the eccentricity of the orbit and $\omega$ is the argument of periastron. Our eccentric model constrains the eccentricity to be $e=0.23\pm0.15$. We compare the eccentric with the circular model by calculating the difference in Bayesian information criterion (BIC, \cite{Schwarz_1978_BIC}) between the models. The BIC is defined as follows:
\begin{equation}\label{equ:BIC}
    BIC = k\ln(n)+\chi^2
\end{equation}
Here, $k$ is the number of parameters fitted, $n$ is the number of data points and $\chi^2$ is the usual fit error. It is somewhat ambiguous how to count the number of data points since each RV data point influences the model much more than each photometric data point. We chose to simply define $n$ as the sum of all RV data points and photometric data points, which pushes $\Delta BIC$ towards the circular model. Despite this bias towards circular model the difference in BIC, given by $\Delta BIC=BIC_{ecc}-BIC_{circ}$ is $-2.9$, indicating positive, but not necessarily conclusive evidence in support of the eccentric model. However, there is a known bias towards measuring slight eccentricity even if the true eccentricity is $0$ \citep{Lucy_1971_bias}. Thus, the parameters listed in Table \ref{tab:hatcur_params} are generated assuming the circular model, with the upper limit on.

Figure \ref{fig:RV_vels} shows the RV data points along with the best fit eccentric model and the $1\sigma$ uncertainty band versus time (upper panel) and versus orbital phase (lower panel), along with the corresponding residuals. The residuals show no significant long term trend, consistent with a single planet Keplerian model. The RV coverage was sufficient to constrain the orbital semi-amplitude to \hatcurRVK\,m\,s$^{-1}$, with a systematic velocity offset of \hatcurRVgamma\,m\,s$^{-1}$. This results in a planet mass of \hatcurPPm\,$M_{Jup}$. The orbital period is consistent with the photometrically determined period at P=\hatcurLCP\, days, firmly placing this planet in the Hot Jupiter regime. All fitted parameters with their uncertainties are shown in Table \ref{tab:hatcur_params}. 

Figure \ref{fig:transit} shows the phase folded TESS light curve, binned into 200 bins, along with the best fit Mandel Agol model in red. We infer a planetary radius of $R_p$ = \hatcurPPr\,$R_{J}$. The according best fit planetary density is $\rho =$\hatcurPPrho$\mathrm{g\,cm^{-3}}$, consistent with an inflated hot Jupiter. The phase folded light curve appears to display a slight rise in out of transit flux near zero phase. This variation is likely an artifact of the TFA detrending. We do not attempt to subtract out this effect here. We caution that the TFA process may have slightly suppressed the true transit depth, producing a smaller planet radius measurement than the true value. Typically TFA-processed light curves are fit using a scaling factor that accounts for this dilution. However, as we do not have any follow-up light curve that can be used to properly fix the radius, we cannot fit for a scaling factor in this case. \citet{T16_Joel} find that the standard (non-reconstructive) TFA process typically reduces the radius by $\sim 10$\%.

\begin{table}[h]
\centering
\caption{Best fit system parameters for \hatcurhtr}
\scriptsize
\label{tab:hatcur_params}
\setlength{\tabcolsep}{0.5pt}
\begin{tabular}{lcc}
\hline\hline
Parameter & Symbol & Value \\
\hline
\multicolumn{3}{c}{\textit{Stellar parameters}} \\
\hline
Right ascension (J2000)             & $\alpha$                 & \hatcurCCra \\
Declination (J2000)                 & $\delta$                 & \hatcurCCdec \\
Effective temperature               & $T_{\rm eff}$ (K)        & \hatcurISOteff \\
Metallicity                         & [Fe/H]                   & \hatcurISOzfeh \\
Stellar mass                        & $M_\star$ ($M_\odot$)    & \hatcurISOm \\
Stellar radius                      & $R_\star$ ($R_\odot$)    & \hatcurISOr \\
Surface gravity                     & $\log g_\star$ (cgs)     & \hatcurISOlogg \\
Mean density                        & $\rho_\star$ (g cm$^{-3}$) & \hatcurISOrho \\
Age                                 & (Gyr)                    & \hatcurISOage \\
Distance                            & $d$ (pc)                 & \hatcurXdistred \\
Extinction                          & $A_V$ (mag)              & \hatcurXAv \\
\hline
\multicolumn{3}{c}{\textit{Photometry}} \\
\hline
Gaia $G$                            & $G$ (mag)                & \hatcurCCgaiamG \\
Gaia $G_{\rm BP}$                   & $G_{\rm BP}$ (mag)       & \hatcurCCgaiamBP \\
Gaia $G_{\rm RP}$                   & $G_{\rm RP}$ (mag)       & \hatcurCCgaiamRP \\
2MASS $J$                           & $J$ (mag)                & \hatcurCCtwomassJmag \\
2MASS $H$                           & $H$ (mag)                & \hatcurCCtwomassHmag \\
2MASS $K_s$                         & $K_s$ (mag)              & \hatcurCCtwomassKmag \\
WISE $W1$                           & $W1$ (mag)               & \hatcurCCWonemag \\
WISE $W2$                           & $W2$ (mag)               & \hatcurCCWtwomag \\
\hline
\multicolumn{3}{c}{\textit{Orbital and transit parameters}} \\
\hline
Orbital period                      & $P$ (days)               & \hatcurLCP \\
Transit epoch                     & $T_0$ (BJD$_{\rm TDB} -2457000$)  & \hatcurLCT \\
Transit duration                    & $T_{14}$ (days)          & \hatcurLCdur \\
Scaled semimajor axis               & $a/R_\star$              & \hatcurPPar \\
Impact parameter                    & $b$                      & \hatcurLCimp \\
Radius ratio                        & $R_p/R_\star$            & \hatcurLCrprstar \\
Eccentricity (upper limit)                        & $e$                      & $<0.506$ \\
RV semi-amplitude                   & $K$ (m s$^{-1}$)         & \hatcurRVK \\
\hline
\multicolumn{3}{c}{\textit{Planetary parameters}} \\
\hline
Planet mass                         & $M_p$ ($M_{\rm Jup}$)    & \hatcurPPm \\
Planet radius\footnote{The planet radius may be underestimated slightly due to the lack of follow-up photometry}                      & $R_p$ ($R_{\rm Jup}$)    & \hatcurPPr \\
Planet density                      & $\rho_p$ (g cm$^{-3}$)   & \hatcurPPrho \\
Surface gravity                     & $\log g_p$ (cgs)         & \hatcurPPlogg \\
Equilibrium temperature             & $T_{\rm eq}$ (K)         & \hatcurPPteff \\
Safronov number                     & $\Theta$                 & \hatcurPPtheta \\
Circularization timescale           & $t_{\rm circ}$ (Myr)     & \hatcurPPtcirc \\
Orbital decay timescale             & $t_{\rm infall}$ (Myr)   & \hatcurPPtinfall \\
\hline
\end{tabular}
\end{table}

\subsection{Blend Analysis}

To further validate the planetary nature of our candidate, we conduct blend analyses of the TESS light curve for our target following the methods of \citet{Hartman_2019_lfit}. Common astrophysical false positives include ($1$) hierarchical triples in which the fainter stars form an eclipsing binary blended with a brighter primary star and ($2$) a chance alignment between an eclipsing binary and an unrelated bright foreground (or background) star. For each blend scenario we first identify the best fit model using a downhill-simplex minimization, which provides the starting point for a subsequent differential evolution Markov Chain Monte Carlo analysis. 

In the hierarchical triple model we allow the transit times of two reference transits, the age of the system, the mass of the three stellar components, the distance of the system, the metallicity of the system, and the impact parameter of the eclipse to vary. For the blend between the foreground star and the eclipsing binary system we vary those same parameters, as well as the age, metallicity, and distance of the blended foreground star. We compare these false positive scenarios to our true positive model, a star-planet system. For the star-planet system we vary two transit reference times, the age of the system, the ratio between the planet and the star radii, the mass of the star and the planet, the distance and metallicity of the system, and the impact parameter of the transit. All stellar parameters in the blend models are constrained to lie on MIST isochrones, ensuring self-consistent masses, radii, luminosities, and temperatures.

To evaluate the goodness of fit we compute the Bayesian Information Criterion (BIC, Equation \ref{equ:BIC}) for each scenario. We find that the transiting planet model is very strongly favored over both the hierarchical triple scenario and the background EB scenario, with $BIC_{planet}-BIC_{ht}=116.4$ and $BIC_{planet}-BIC_{bgEB}=27.9$, where $BIC_{planet}$, $BIC_{ht}$, and $BIC_{bgEB}$ refer to the BIC values for the planetary model, the hierarchical triple model, and the background EB model respectively. This analysis thus allows us to confidently rule out these false positive scenarios, further confirming the planetary nature of the signal.

\subsection{Connection to current hot Jupiter formation theory}

Figure \ref{fig:isochrones} shows the position of the target on a stellar density versus effective temperature plot, along with its $1\sigma$ uncertainty region. Plotted in various colors are stellar isochrones, where the times are in Gyrs. The target's position near the turnoff of the isochrones indicates that the host star is old and likely slightly evolved, a conclusion that is also supported by the kinematic association of the star with the Galactic thick disk. Using the tidal circularization timescale formula given by \cite{Jackson_2008_tidal} in equation 4 we can estimate a characteristic timescale for this planet:
\begin{equation}
    \tau_{circ}\approx\frac{4Q_pM_p}{63\sqrt{GM_*^3}R_p^{5}}a^{13/2}
\end{equation}
Where G is the gravitational constant, $Q_p$ is the tidal quality factor, which we assume to be $\sim10^6$, $a$ is the semimajor axis, $R_p$ is the planetary radius and $M_p$ and $M_*$ are the 
planetary and stellar mass, respectively. Here, we ignore the circularizing effects of tides raised on the star by the planet as this constitutes a higher order process and assume a constant semi-major axis. Plugging in our best fit values we get 
a tidal circularization timescale of $t_e\sim 0.5$~Gyrs. The preference for slight eccentricity is marginal and could thus be an artifact of our fitting procedure. Equation $13$ in \citet{Hara_2019} allows the quantification of this so-called Lucy-Sweeney bias \citep{Lucy_1971_bias} if the true eccentricity is assumed to be near zero:
\begin{equation}
    b=\sqrt{\frac{\pi}{4-\pi}}\sigma_e
\end{equation}
where $\sigma_e$ is the measured uncertainty on the eccentricity. We measured a uncertainty of $\sigma_e=0.15$, resulting in a bias of $b\approx0.287$. This result indicates that our fit is entirely compatible with a circular orbit.
In addition, the presence of potential additional planets in the system could drive eccentricity. If real, however, the slight eccentricity in addition to the old age of the host star could indicate that this planet did not form in situ and instead belongs to the ``late-arrivers'' as described in \cite{Chen_2025_HotJupOrigin}.

Higher precision RV measurements would be required to constrain the planetary parameters better. However, our dataset clearly proves the presence of a previously undiscovered, inflated, hot Jupiter around TIC $183374187$. TIC $183374187$ is the only target from our new candidate list that has been followed up up so far, albeit a promising one based on the light curve properties. While this discovery proves our pipeline's capability of detecting new planetary systems, it does not constrain the false positive rate or the detection sensitivity, which are beyond the scope of this work.

\begin{deluxetable}{cccc}
\tablecaption{
Description of parameters used in the candidate catalog. All quoted uncertainties are at the $1\sigma$ level.
Quantities with uncertainties are derived using the MCMC process described in Section~\ref{Error_estimation}.
Stellar parameters are taken primarily from \textit{Gaia}. The signal-to-pink-noise (S/PN), number of transits,
$q_{\rm tran}$, and $q_{\rm ingress}$ values are derived using our BLS process.
The full candidate catalog is available in the electronic edition of the journal.
\label{tab:parameter_descripton}
}
\tablehead{
\colhead{Row} & \colhead{Units} & \colhead{Parameter} & \colhead{Description}
}
\startdata
1  & \nodata        & GAIAID        & Gaia DR3 source identifier \\
2  & \nodata        & TICID         & TESS Input Catalog identifier \\
3  & \nodata        & Sector        & TESS observing sector in which the star was observed \\
4  & \nodata        & Cam           & TESS camera number (1--4) \\
5  & \nodata        & CCD           & TESS CCD number (1--4) \\
6  & K              & Teff          & Stellar effective temperature \\
7  & $L_\odot$      & Lum           & Stellar luminosity \\
8  & $R_\odot$      & Rs            & Stellar radius \\
9  & $M_\odot$      & Ms            & Stellar mass \\
10 & mag            & TESS\_mag     & TESS-band magnitude \\
11 & dex            & mh\_gspphot   & Stellar metallicity [M/H] from \textit{Gaia} GSP-Phot \\
12 & \nodata        & qtran         & Transit duration divided by orbital period \\
13 & \nodata        & qingress      & Ingress/egress duration divided by transit duration \\
14 & days           & P\_med        & Median orbital period \\
15 & days           & P\_errm       & Lower (negative) uncertainty on orbital period \\
16 & days           & P\_errp       & Upper (positive) uncertainty on orbital period \\
17 & BJD            & Tc\_med       & Median transit mid-time (BJD-2457000)\\
18 & days           & Tc\_errm      & Lower uncertainty on transit mid-time \\
19 & days           & Tc\_errp      & Upper uncertainty on transit mid-time \\
20 & \nodata        & a\_rs\_med    & Median scaled semi-major axis $a/R_\star$ \\
21 & \nodata        & a\_rs\_errm   & Lower uncertainty on $a/R_\star$ \\
22 & \nodata        & a\_rs\_errp   & Upper uncertainty on $a/R_\star$ \\
23 & \nodata        & rp\_rs\_med   & Median planet-to-star radius ratio $R_p/R_\star$ \\
24 & \nodata        & rp\_rs\_errm  & Lower uncertainty on $R_p/R_\star$ \\
25 & \nodata        & rp\_rs\_errp  & Upper uncertainty on $R_p/R_\star$ \\
26 & \nodata        & b\_med        & Median impact parameter \\
27 & \nodata        & b\_errm       & Lower uncertainty on impact parameter \\
28 & \nodata        & b\_errp       & Upper uncertainty on impact parameter \\
29 & \nodata        & q1\_med       & Median limb-darkening parameter $q_1$ \\
30 & \nodata        & q1\_errm      & Lower uncertainty on $q_1$ \\
31 & \nodata        & q1\_errp      & Upper uncertainty on $q_1$ \\
32 & \nodata        & q2\_med       & Median limb-darkening parameter $q_2$ \\
33 & \nodata        & q2\_errm      & Lower uncertainty on $q_2$ \\
34 & \nodata        & q2\_errp      & Upper uncertainty on $q_2$ \\
35 & deg            & ra            & Right ascension (J2000) \\
36 & deg            & dec           & Declination (J2000) \\
37 & \nodata        & S/PN          & Signal-to-pink-noise detection statistic \\
38 & \nodata        & Ntransits     & Number of observed transits in the TESS light curve \\
\enddata
\tablecomments{
This table is published in its entirety in the electronic edition of the journal.
A portion is shown here for guidance regarding its form and content.
}
\end{deluxetable}

\section{Discussion} \label{sec:discussion}
\begin{figure}
    \centering
    \includegraphics[width=\linewidth]{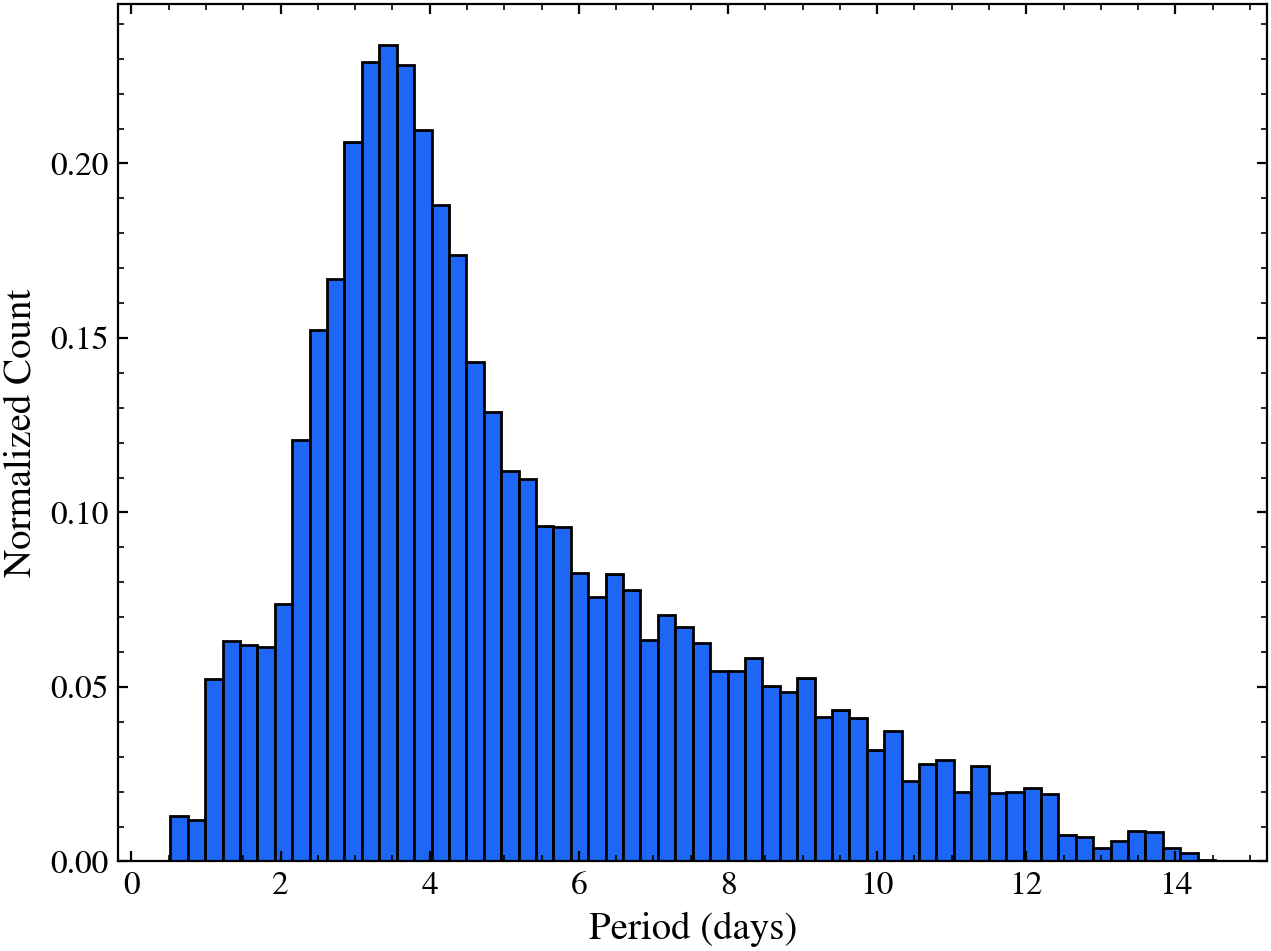}
    \caption{Period distribution of our candidates. The prominent peak around $3.5$ days is expected from Hot Jupiter surveys since Hot Jupiters constitute a majority of our candidates, while the falloff towards longer periods is due to the decreasing transit probability and lower SNR of longer period candidates.}
    \label{fig:Candidate_periods}
\end{figure}

\begin{figure*}[!t]
    \centering
    \includegraphics[width=0.85\linewidth]{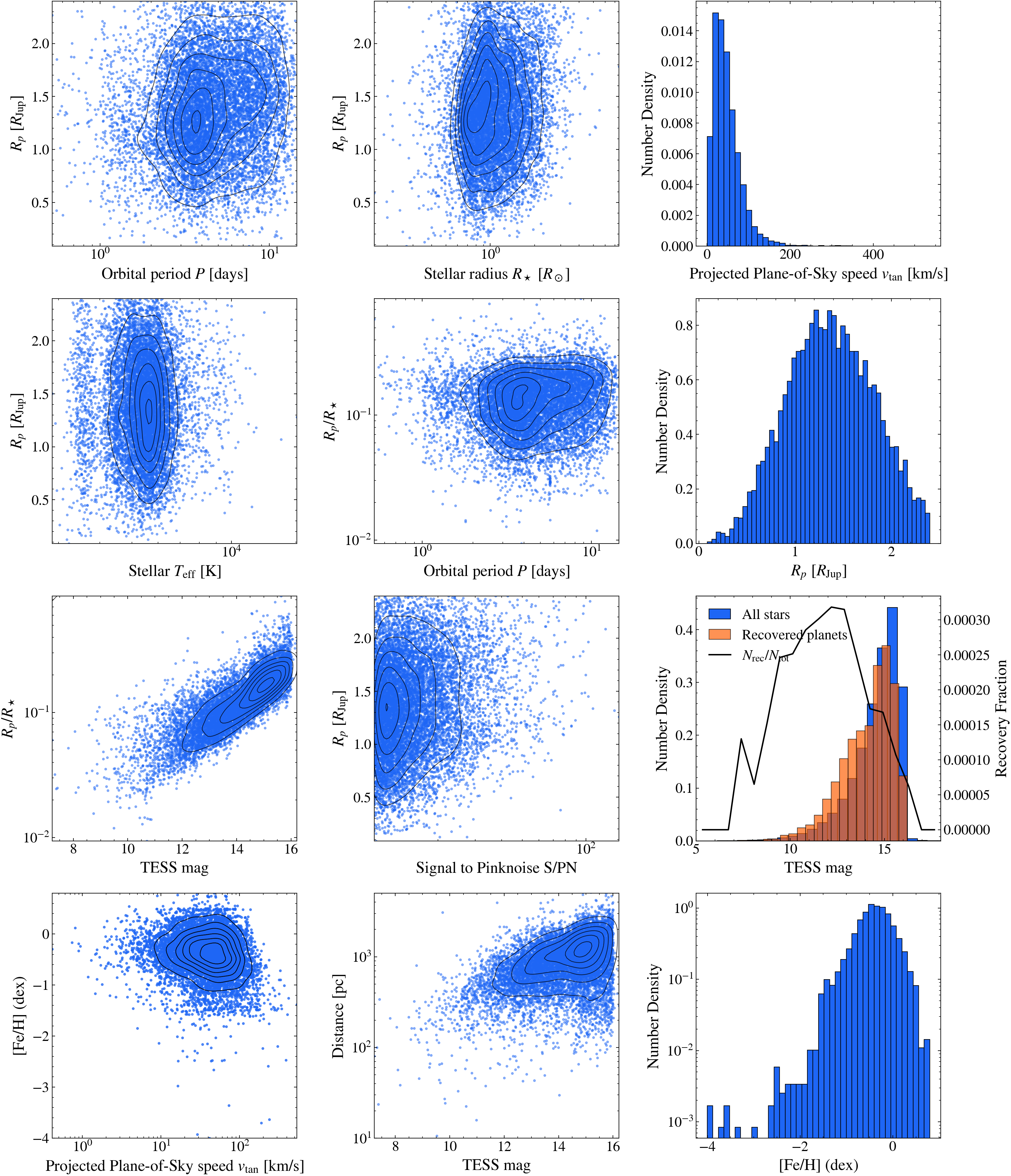}
    \caption{Selected parameters distributions and correlations from our final sample. The orange magnitude histogram shows the TESS magnitude distribution of our candidates, as opposed to the blue histogram, which shows the magnitude distribution of all Cycle $1$ samples analyzed. The black curve shows the fraction of all stars in a magnitude bin that we recovered as planet candidates. Black contours on the scatter plots indicate the smoothed two-dimensional number density of points, estimated using a Gaussian kernel density estimator. Contours are shown at fixed percentile levels of the density distribution to guide the eye. Orange points in the Mollweide projection represent known TOIs/CTOIs recovered by this search, blue points indicate galactic positions of all candidates identified in this work.}
    \label{fig:parameter_plot}
\end{figure*}
Figure \ref{fig:Candidate_periods} shows the orbital period distribution for our candidates. The distribution shows a clear peak around $3.5$ days, which aligns neatly with the canonical ``pile-up'' of hot Jupiter orbital periods between $3$ and $5$ days first identified in RV-surveys such as \cite{Cumming_2008_period_peak}. While the large majority of the candidates would be gas-giant-size planets, the sample also includes smaller planet candidates as well (see below). This peak is encouraging, but our candidate sample of course does not reflect the true period distributions for hot Jupiters. For example, longer period signals are suppressed due to their lower SNR, which biases the RFC to misclassify such candidates, and by the fact that transit probability decreases as $P^{-2/3}$, intrinsically suppressing the number of long-period transits. TESS's $27$-day observing baseline per sector further reduces detectability beyond $P\gtrsim13$ days for single-sector targets and contributes to the deficit of longer-period systems in our catalog. This underlines the necessity to combine multiple TESS sectors in future works to increase the sensitivity to longer period planets.

Figure \ref{fig:parameter_plot} shows selected parameter distributions and correlations from our final catalog. Of our $11,143$ candidates, $3,338$ are brighter than $T=13.5$ and therefore fall within the magnitude range searched by SPOC, QLP, or other TESS pipelines. The vast majority of these bright candidates are listed as TESS threshold crossing events (TCEs, https://tev.mit.edu/), yet $2,479$ are not promoted to TOIs, despite the absence of explicit fail flags for most of them. While the detections of these bright candidates appear robust based on our vetting plots, there is some possibility that information from subsequent TESS cycles or follow up observations, neither of which are considered here, led to the non-selection of the targets by these other surveys. 

\begin{deluxetable}{ccc}
\tablecaption{
Description of parameters used in the single transit candidate catalog. 
The full candidate catalog is available in the electronic edition of the journal. Parameter values are derived from the BLS fits and represent approximate estimates based on the detection pipeline; they should therefore be interpreted with caution. 
\label{tab:singles_parameter_descripton}
}
\tablehead{
\colhead{Row} & \colhead{Parameter} & \colhead{Description}
}
\startdata
1        & GAIAID        & Gaia DR3 source identifier \\
2     & TICID         & TESS Input Catalog identifier \\
3      & Sector        & TESS observing sector in which the star was observed \\
4      & Cam           & TESS camera number (1--4) \\
5      & CCD           & TESS CCD number (1--4) \\
6       & Duration  & Transit duration (days) \\
7       & Tc        & Transit mid-time (BJD-$2457000$)\\
8       & Depth     & Transit depth (mag)\\
\enddata
\tablecomments{
This table is published in its entirety in the electronic edition of the journal.
A portion is shown here for guidance regarding its form and content.
}
\end{deluxetable}

\begin{figure}
    \centering
    \includegraphics[width=\linewidth]{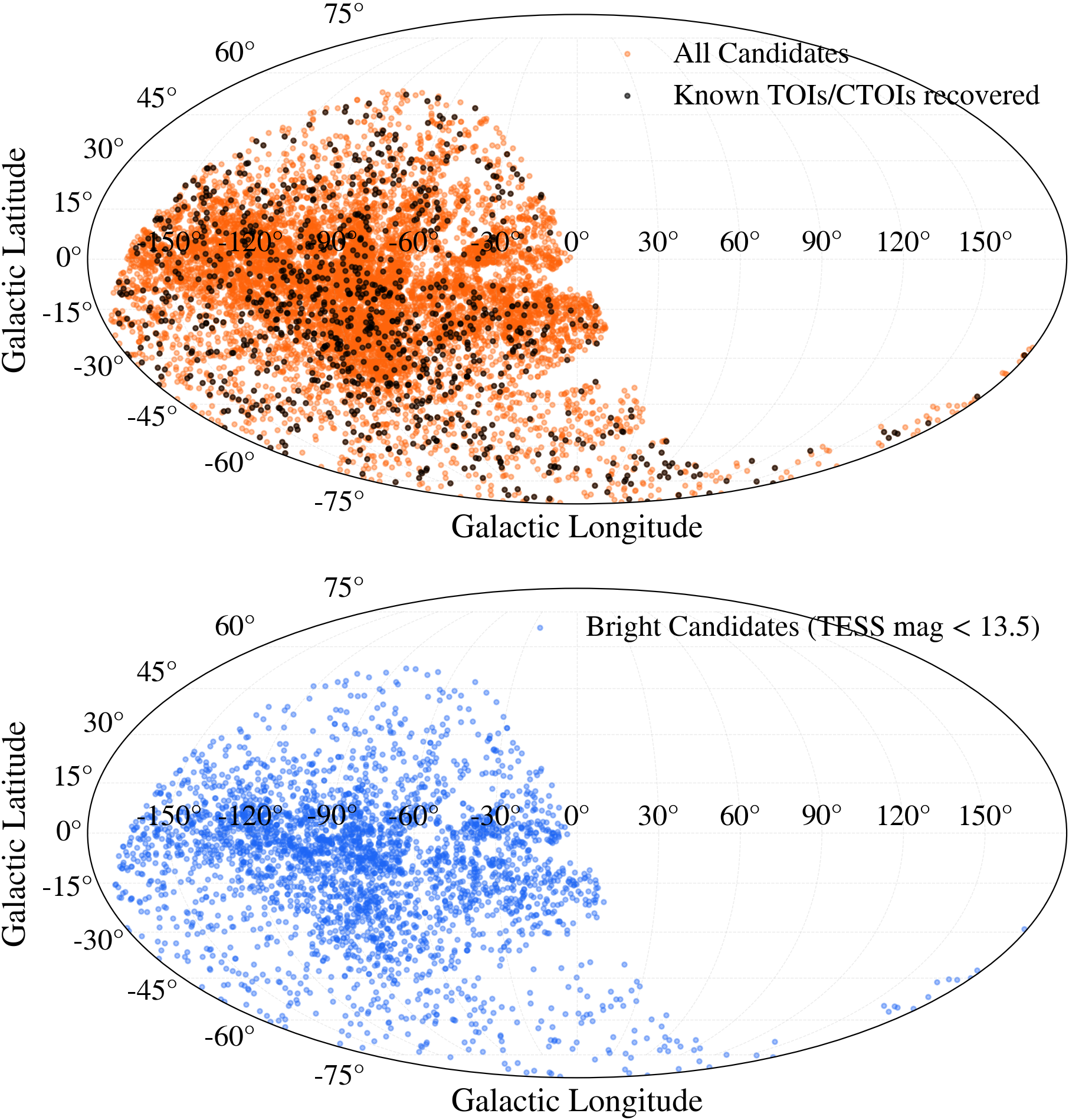}
    \caption{Distribution of Galactic coordinates for our entire candidate set (top panel, orange data points), our recovered TOIs/CTOIs (top panel, black data points), and our $T<13.5$\,mag candidates (lower panel, blue data points).}
    \label{fig:Mollweide}
\end{figure}

Figure \ref{fig:Mollweide} shows the distribution of Galactic coordinates for recovered TOIs and CTOIs (top panel, black data points), all our candidates (top panel, orange data points), and our bright candidates (lower panel, blue data points). Our bright candidates exhibit a modestly stronger concentration toward the Galactic bulge compared to the previosuly known candidates recovered here. This trend may reflect the effectiveness of our image-subtraction-based light curve generation in highly crowded fields, where blending can limit the sensitivity of other pipelines, enabling the identification of candidates that were missed in prior searches. The vast majority ($70\%$) of our candidates are fainter than this however, highlighting the benefit of conducting deeper transit searches. 

Planet classification typically requires some knowledge about planetary composition, but we can roughly classify our candidates into categories using the inferred radii. We adopt approximate exoplanet classifications of $1R_\oplus \le R_p< 2R_\oplus$ for super-Earths, $2R_\oplus \le R_p < 3.9R_\oplus$ for sub-Neptunes, $3.9R_\oplus \le R_p < 0.5R_{J}$ for Neptunians, and $0.5R_J\le R_p$ for Gas-Giants \citep{Chen_2016_classifications, Rogers_2015_Superearths} and find that most of our candidates ($97.7\%$) are Gas-Giants, followed by Neptunians ($1.5\%$) and sub-Neptunes ($0.64\%$). We find $11$ potential super-Earths. 

Our candidate list contains $737$ low metallicity stars with [Fe/H]$<-1$ based on \textit{Gaia} DR3 stellar parameters, which could belong to older populations in the Galactic thick disk or even the halo. We furthermore find $66$ ultra-short period (USP) candidates with orbital periods less than a day.

We do not attempt to constrain the orbital periods of the mono-transit candidates in our sample in any way, and even the transit times may be inaccurate. Instead, our sample of mono transit candidates should be viewed simply as a collection of threshold crossing single transit signals, which can be used in future work as a starting point for cross sector searches for longer warm or cold Jupiters. 

\subsection{Recovered TOIs}\label{sec:TOI}

It is a useful check of our pipeline to explore its capacity to recover known TESS objects of interest (TOIs). Our search returned $928$ TOIs. Of these, $826$ are listed in the TOI catalog as having at least one Cycle 1 light curve. The other $102$ do not have a Cycle 1 light curve listed in the TOI catalog but we have provided one as part of the T$16$ project. We call the former ``cycle $1$ TOIs'', although many  were formally identified only after additional sectors were incorporated into the broader TOI vetting process. $169$ of the $928$ TOIs we recover have been confirmed as planets, while $76$ have been identified as false positives. The current TOI catalog includes $3,549$ such Cycle $1$ TOIs around $3400$ targets.

Our CETRA transit recovery recovers $2095$ cycle $1$ TOIs (within $1\%$ of their reported orbital periods), and for only $1629$ of these we find $S/PN>10$, making them recoverable by our pipeline. Our pipeline misses $766$ of these recoverable TOIs, of which $166$ have been identified as false positives, and $86$ have been confirmed as real planet signals to date. Our RFC, however, classifies $709$ of these TOIs ($93\%$) as planet candidates. Of the $709$ TOIs that pass our RFC $72$ are eliminated in the manual vetting process, the other $637$ are eliminated in roughly equal amounts by the contamination testing, the period bin clipping, and the radius estimation routines described in Section \ref{Sec:automated_vetting}. Although this vetting process reduces the recovery rate of known TOIs, it is essential to establish confidence in the much larger sample of newly identified candidates. Future work may explore less conservative versions of these cuts to further improve completeness.

\subsection{Stellar Parameters of Catalog hosts}

\begin{figure}
    \centering
    \includegraphics[width=\linewidth]{HR_diagram.png}
    \caption{Distribution of our candidates on an HR diagram. The red cross marks the Sun's position, the background colors provide approximate stellar types via the classifications in \citet{Pecaut_2013_stellarTypes}. Stellar parameters are drawn from the TESS Input Catalog \citep{Stassun_2018_TIC}, supplemented by \textit{Gaia} DR3 where necessary \citep{Gaia_DR3}}.
    \label{fig:HR_diagram}
\end{figure}

Figure \ref{fig:HR_diagram} shows the distribution of hosts on a Hertzsprung-Russell (HR) diagram, with background colors roughly indicating the stellar group. This allows follow up work to be concentrated on candidates that fall within a particular interesting region of the HR diagram, such as late M-dwarfs. These interesting candidates, along with other targets, such as candidates around metal poor stars that could lie in the Galactic halo, are being followed up and will be discussed in more detail in subsequent work. 

Using these stellar parameters furthermore allows us to categorize our candidate sources into stellar classes. We find that $5.2\%$ of our candidates orbit M-dwarfs, $37.3\%$ orbit K-type stars, $33.5\%$ orbit G-type stars, $18.2\%$ orbit F-types stars, $3.1\%$ orbit A-type stars and only $0.38\%$ of our candidates orbit B-type stars.

\subsection{Future Work/Follow ups}
As previously mentioned, a natural extension of this project is a cross-sector search designed to identify longer-period candidates. Such an effort will benefit from the inclusion of additional TESS cycles. The generation of T$16$ light curves for Cycles 2 and 3 is already well underway, and we have initiated the transit search for these extended datasets. Incorporating these cycles will significantly improve our sensitivity to multi-sector events and will likely add thousands of additional candidates to our catalog.

Many of the newly discovered candidates are relatively faint and would require substantial telescope time for radial velocity or high-resolution imaging follow-up. A practical strategy is therefore to prioritize the most astrophysically informative targets. Our catalog contains a large number of late M-dwarf hosts, offering opportunities to study the formation and survival of short-period gas giants around low-mass stars. Additionally, the presence of hot Jupiter candidates orbiting metal-poor stars enables tests of the metallicity dependence of giant-planet formation, a topic for which we are actively collecting follow-up observations.

Although follow-up observations of faint stars are difficult, it is possible for stars down to $16^\mathrm{th}$ magnitude (and beyond) if the scientific interest warrants the telescope time necessary. One can estimate the telescope time required to follow up various planet types around $16^\mathrm{th}$ magnitude stars by first noting that the semi-amplitude of the RV signal is about
\begin{equation}\label{eq:K}
    K = 28.4\,\mathrm{ms^{-1}}\left(\frac{M_p}{M_J}\right)\left(\frac{1yr}{P_p}\right)^{1/3}\left(\frac{M_\odot}{M_s}\right)^{2/3}
    \sin(i)
\end{equation}
where $M_p$ and $P_p$ are the mass and orbital period of the planet, $M_s$ is the mass of the host star and $i$ is the inclination. For this estimate we assume $i=90\degree$ and that the star is of solar mass. Equation \ref{eq:K} is a direct result from the binary mass function in the limit $M_p<<M_s$. To get a reliable confirmation of a planetary signal at $5\sigma$ level the RV uncertainty $\sigma_{RV}$ has to be $\le \frac{K}{5}$. The RV uncertainty scales as $\sigma_{RV}\propto\frac{1}{\sqrt{N}}$, where N is the number of photons received. Thus, N is proportional to the received flux and the exposure time $t$. Rewritten in magnitude space this gives the relation:
\begin{equation}\label{eq:expTime}
    t=t_0\left(\frac{\sigma_0}{\sigma_{RV}}\right)^210^{0.4(m-m_0)}
\end{equation}
Here, $m$ is the magnitude of the target star, and $\sigma_0$, $t_0$, and $m_0$ are the RV uncertainty, exposure time, and magnitude of some reference star, for which we choose TIC$183374187$ ($m_0=14.09$, $t_0=1200\,\mathrm{s}$, $\sigma_0\approx12\,\mathrm{ms^{-1}}$). Combining equations \ref{eq:K} and \ref{eq:expTime} under the assumption that we need $\sigma_{RV}\le \frac{K}{5}$, we get a formula to obtain an informative RV data point using PFS:
\begin{equation}\label{eq:scaling}
t \approx
\frac{5356\,\mathrm{s}\times 10^{0.4(m-14.09)}}
{
\sin^2(i)
\left(\frac{M_p}{M_J}\right)^2
\left(\frac{1\,\mathrm{yr}}{P_p}\right)^{2/3}
\left(\frac{M_\odot}{M_s}\right)^{4/3}
}.
\end{equation}
Using this equation, one can estimate the exposure time required for a variety of planet categories. For instance, to get an RV data point for a $1M_J$ hot Jupiter on a $10$ day orbit around a $16^\mathrm{th}$ magnitude Sun like star (assuming $i=90\degree$) one would need an exposure time of about $2830\,\mathrm{s}=0.8\,\mathrm{h}$. The confirmation of a $10\,\mathrm{M_\oplus}$ super-Earth around such a star would require $2.9\times10^6\,\mathrm{s}\approx820\,\mathrm{h}$. At even smaller planetary radii the limiting factor is the RV precision. In general, modern spectrographs struggle to reach RV uncertainties below about $1\,\mathrm{ms^{-1}}$. In practice, this total exposure time would be divided into multiple observations to detect the relative RV variation and sample the orbital phase. We note that the scaling given in Equation \ref{eq:scaling} has been normalized using PFS observations and may differ if other spectrographs are used.

Alternatively, obtaining two RV points with a precision of $\sim1\,\mathrm{km}\,\mathrm{s^{-1}}$ is often sufficient to rule out eclipsing binary false positives. Such a measurement would only require $2\,\mathrm{s}$ of exposure time for the hot Jupiter case described above. In practice, however, instrumental overheads mean that significantly longer would be taken. Ignoring these caveats for now, we can calculate the total exposure time required to follow up all our $11,143$ (multi-transit) candidates in this way:
\begin{equation}
    t_{tot} \approx 0.35\,\mathrm{s}\times \sum_i10^{0.4(m_i-14.09\,\mathrm{mag})}\approx40,000\,\mathrm{s}\approx11\,\mathrm{h}
\end{equation}
Here, $m_i$ are the TESS magnitudes of our multi-transit candidates. This estimate is a strict lower bound, but it demonstrates the feasibility of such a follow-up effort, which would greatly increase the confidence in the candidates presented in this work.

Beyond these astrophysical directions, several methodological improvements could refine our catalog. First, the vetting process stands to benefit from more sophisticated machine-learning models. Incorporating convolutional neural networks (CNNs) to classify full-frame-image (FFI) cutouts, in conjunction with the random forest classifier applied to the light curves, may reduce or even eliminate the need for manual vetting. Our current sample, which includes both reliable positives and well-characterized false positives, provides an excellent training set for such future models.

Furthermore, detailed completeness and reliability analyses such as injection-recovery tests performed on T$16$ light curves in combination with extensive follow-up work to determine the astrophysical false positive rate would enable the use of this catalog for robust occurrence rate studies. As additional TESS data become available, these tests will allow us to place the catalog on a firmer statistical foundation and extend its scientific utility well beyond the initial transit search. However, we note that a full estimate of the completeness of this catalog via injection-recovery tests would be computationally expensive, as the transit search would need to be repeated multiple times for each light curve. Such an effort would likely require substantially more compute time than the entire transit search described in this work, which lasted approximately six months. Nevertheless, this type of analysis may be feasible with modern GPU servers, which can be roughly $50$ times faster than the hardware used for this search. While a comprehensive completeness analysis is beyond the scope of the present work, it presents a promising avenue for future follow-up.

Another possible approach to assessing the reliability of the candidates presented in this work is the use of statistical validation tools such as \texttt{TRICERATOPS} \citep{Giacalone_2021_TRICERATOPS}, which estimates the probability that a detected transit signal is of planetary origin based on the local stellar background and a set of astrophysical false-positive scenarios. However, caution is warranted when applying such methods to hot Jupiter candidates. In practice, \texttt{TRICERATOPS} often returns higher false-positive probabilities for large-radius planets on short orbital periods, where transit signals can be degenerate with eclipsing binary configurations. This may be particularly relevant for samples dominated by hot Jupiter candidates, such as the one considered here.

\section{Conclusions} \label{sec:conclusions}

In this work we analyzed $83,717,159$ detrended TESS FFI light curves produced as part of our T$16$ project. We applied the CETRA transit recovery algorithm and extracted preliminary transit parameters using a BLS search, which we classified using random forest classifiers. Light curves classified as transiting planet candidates were subsequently passed through a suite of automated vetting procedures, including cross-matching with nearby contaminating sources with similar ephemerides, fitting a transit model and removing candidates with unphysical inferred planetary radii, and checks for off-center variability. The resulting set of candidates was then manually reviewed, although it is important to note that due to the large number of candidates that needed to be inspected, a very thorough inspection of each candidate is not feasible. As such, there may be a higher than typical number of eclipsing binaries and other non-planetary systems that are included in the candidate list. We strongly suggest that the vetting plots be inspected for any candidates before performing follow-up observations on them. Our final list of $11, 554$ planet candidates consists of $10,502$ new planet candidates and $1,052$ previously identified candidates. The new candidates include $10,091$ objects with multiple transits detected, for which we report full transit parameters and uncertainties, and $411$ single-transit candidates, for which we report the TESS sector, camera, and CCD of detection. Both catalogs are provided as online supplement, and the primary catalog will be published as CTOIs. This work increases the total number of candidates identified from TESS observations by a factor of $2.4$, while only relying on Cycle 1 observations. It represents the largest homogeneous search for planets in TESS FFIs to date.

We also present follow-up RV observations of one of our candidates, TIC $183374187$, which confirms its planetary nature. Joint modeling of TESS photometry and PFS RVs reveals a $0.58M_J$ hot Jupiter on a $5.059$ day orbit around an old, metal poor host star. The star is determined to be a kinematic member of the Galactic thick disk, and also exhibits alpha-element enhancement consistent with this conclusion. We confidently rule out false positive scenarios such as eclipsing binaries, hierarchical triples, or background blended binaries. 

This transit search is based on $\sim 15\%$ of currently available data. Thus, extending the search to include additional TESS cycles and implementing a cross sector transit search will substantially increase the number of candidates and improve completeness, especially for medium to longer period transits and lower radius candidates. Injection-recovery tests may further characterize the efficiency of the search pipeline, enabling demographic studies of the planet population detected in TESS FFIs and the identification of rare systems well suited for targeted follow up. 

\section*{Acknowledgments}
We thank the anonymous referee for their comments, which greatly improved the quality of this paper.
J.R. acknowledges helpful discussions with Caleb Lammers, Matthew Goodbred, and Boqin Zhang. The authors thank Dr$.$ Veselin Kostov for sharing the TESS $10,000$ EB Catalog before it was publicly available. This
project received funding from the NASA Astrophysics Data
Analysis Program (ADAP) grant 80NSSC22K0409, and from the NASA Exoplanet Research Program (XRP) grant 80NSSC22K0315. This paper includes data collected with the TESS mission, obtained from the MAST data archive at the Space Telescope Science Institute (STScI). Funding for the TESS mission is provided by the NASA Explorer Program. STScI is operated by the Association of Universities for Research in Astronomy, Inc., under NASA contract NAS 5–26555. We acknowledge the use of public TESS data from pipelines at the TESS Science Office and at the TESS Science Processing Operations Center. Resources supporting this work were provided by the NASA High-End Computing (HEC) Program through the NASA Advanced Supercomputing (NAS) Division at Ames Research Center for the production of the SPOC data products. This paper includes data gathered with the 6.5 meter Magellan Telescopes located at Las Campanas Observatory, Chile. This research has made use of the NASA Exoplanet Archive, which is operated by the California Institute of Technology, under contract with the National Aeronautics and Space Administration under the Exoplanet Exploration Program. This work has made use of results from the European Space Agency (ESA) space mission Gaia. Gaia data are being processed by the Gaia Data Processing and Analysis Consortium (DPAC). Funding for the DPAC is provided by national institutions, in particular the institutions participating in the Gaia MultiLateral Agreement (MLA). The Gaia mission website is https://www.cosmos.esa.int/gaia. The Gaia archive website is https://archives.esac.esa.int/gaia. This research has made use of the VizieR catalogue access tool, CDS, Strasbourg, France. This publication makes use of data products from the Two Micron All Sky Survey, which is a joint project of the University of Massachusetts and the Infrared Processing and Analysis Center/California Institute of Technology, funded by the National Aeronautics and Space Administration and the National Science Foundation. This publication makes use of data products from the Wide-field Infrared Survey Explorer, which is a joint project of the University of California, Los Angeles, and the Jet Propulsion Laboratory/California Institute of Technology, funded by the National Aeronautics and Space Administration.
\\
\\
\textbf{\textit{Facility:}} \textit{Gaia} \citep{GAIA_2016, Gaia_DR3}, Magellan PFS \citep{Crane_2006_PFS, Crane_2008_PFS, PFS_Magellan}, TESS \citep{TESS}
\\
\\
\textbf{\textit{Software:}} \texttt{astropy} \citep{astropy}, \texttt{astroquery} \citep{astroquery}, \texttt{BATMAN} \citep{BATMAN_Kreidberg_2015}, \texttt{DAOStarFinder} \citep{DAOStarFinder}, \texttt{emcee} \citep{Foreman-Mackey_2013_emcee}, \texttt{jktebop} \citep{Southworth_2008_jktebop}, \texttt{matplotlib} \citep{Hunter_2007_matplotlib}, \texttt{Networkx} \citep{networkx}, \texttt{numpy} \citep{Harris_2020_numpy}, \texttt{pandas} \citep{mckinney2010_pandas}, \texttt{PyAstronomy} \citep{pyastronomy}, \texttt{scikit-learn} \citep{scikit-learn}, \texttt{VARTOOLS} \citep{VARTOOLS}
\clearpage
\appendix
\begin{center}
    
\section{Base model RFC parameters}
\begin{deluxetable*}{ll}[h]
\tablecaption{Features used to train the base random forest classifier (RFC).\label{tab:rf_features_base}}
\tablehead{
\colhead{Parameter} & \colhead{Description}
}
\startdata
BLSFixPer\_SignaltoPinknoise\_0 & Signal-to-pink-noise ratio of the primary CETRA peak \\
BLSFixPer\_SignaltoPinknoise\_1 & Signal-to-pink-noise ratio at half the period returned by CETRA\\
BLSFixPer\_SignaltoPinknoise\_2 & Signal-to-pink-noise ratio at twice the period returned by CETRA\\
BLSFixPer\_SignaltoPinknoise\_3 & Signal-to-pink-noise ratio at three times the period returned by CETRA \\
BLSFixPer\_SR\_0 & Signal residue (SR) of the primary CETRA peak \\
BLSFixPer\_SR\_1 & Signal residue (SR) at half the period returned by CETRA \\
BLSFixPer\_SR\_2 & Signal residue (SR) at twice the period returned by CETRA \\
BLSFixPer\_SR\_3 & Signal residue (SR) at three times the period returned by CETRA \\
BLSFixPer\_deltaChi2\_0 & $\Delta\chi^2$ improvement for the primary CETRA peak \\
BLSFixPer\_deltaChi2\_1 & $\Delta\chi^2$ improvement at half the period returned by CETRA \\
BLSFixPer\_deltaChi2\_2 & $\Delta\chi^2$ improvement at twice the period returned by CETRA \\
BLSFixPer\_deltaChi2\_3 & $\Delta\chi^2$ improvement at three times the period returned by CETRA \\
BLSFixPer\_deltaChi2\_invtransit\_0 & $\Delta\chi^2$ for inverted primary transit model \\
BLSFixPer\_deltaChi2\_invtransit\_1 & $\Delta\chi^2$ for inverted transit model at half the period returned by CETRA \\
BLSFixPer\_deltaChi2\_invtransit\_2 & $\Delta\chi^2$ for inverted transit model at twice the period returned by CETRA \\
BLSFixPer\_Depth\_0 & Transit depth of the primary CETRA signal \\
BLSFixPer\_Depth\_1 & Transit depth at half the period returned by CETRA \\
BLSFixPer\_Depth\_2 & Transit depth at twice the period returned by CETRA\\
BLSFixPer\_Depth\_3 & Transit depth at three times the period returned by CETRA \\
BLSFixPer\_Qtran\_0 & Fractional transit duration ($T_{\rm dur}/P$) for the primary signal \\
BLSFixPer\_Qtran\_1 & Fractional transit duration ($T_{\rm dur}/P$) at half the period returned by CETRA \\
BLSFixPer\_Npointsintransit\_0 & Number of in-transit data points for the primary signal \\
BLSFixPer\_fraconenight\_0 & Fraction of in-transit points from a single transit (primary) \\
BLSFixPer\_fraconenight\_1 & Fraction of in-transit points from a single transit (at half the period returned by CETRA) \\
BLS\_HarmAmp\_0 & Amplitude of the strongest harmonic variability component \\
BLS\_HarmMean\_0 & Mean level of the strongest harmonic variability component \\
BLS\_HarmDeltaChi2\_0 & $\Delta\chi^2$ improvement from the harmonic variability model \\
BLS\_harmA\_0 & Harmonic model coefficient $A$ \\
Teff & Stellar effective temperature \\
lum & Stellar luminosity \\
\enddata

\tablecomments{
BLS-based parameters are derived from fixed-period BLS searches and their harmonics as well as a narrow period band BLS call. Stellar parameters provide host-star context during classification.}
\end{deluxetable*}

\clearpage
\FloatBarrier
\mbox{} 

\newpage
\newpage
\section{Faint star RFC parameters}

\begin{deluxetable*}{ll}[h!]
\tablecaption{Features used to train the random forest classifiers for faint stars.\label{tab:rf_features}}
\tablehead{
\colhead{Parameter} & \colhead{Description}
}
\startdata
BLSFixPer\_SignaltoPinknoise\_0 & Signal-to-pink-noise ratio of the primary CETRA peak \\
BLSFixPer\_SignaltoPinknoise\_1 & Signal-to-pink-noise ratio at half the period returned by CETRA \\
BLSFixPer\_SignaltoPinknoise\_2 & Signal-to-pink-noise ratio at twice the period returned by CETRA \\
BLSFixPer\_SR\_0 & Signal residue (SR) of the primary CETRA peak \\
BLSFixPer\_SR\_1 & Signal residue (SR) at half the period returned by CETRA \\
BLSFixPer\_SR\_2 & Signal residue (SR) at twice the period returned by CETRA \\
BLSFixPer\_SR\_3 & Signal residue (SR) at three times the period returned by CETRA \\
BLSFixPer\_deltaChi2\_0 & $\Delta\chi^2$ improvement for the primary CETRA transit model \\
BLSFixPer\_deltaChi2\_1 & $\Delta\chi^2$ improvement for the transit model at half the period returned by CETRA \\
BLSFixPer\_deltaChi2\_2 & $\Delta\chi^2$ improvement for the transit model at twice the period returned by CETRA \\
BLSFixPer\_deltaChi2\_3 & $\Delta\chi^2$ improvement for the transit model at half the period returned by CETRA \\
BLSFixPer\_deltaChi2\_invtransit\_0 & $\Delta\chi^2$ for inverted primary transit model \\
BLSFixPer\_deltaChi2\_invtransit\_1 & $\Delta\chi^2$ for inverted transit model at half the period returned by CETRA \\
BLSFixPer\_deltaChi2\_invtransit\_2 & $\Delta\chi^2$ for inverted transit model at twice the period returned by CETRA \\
BLSFixPer\_Depth\_0 & Transit depth of the primary CETRA signal \\
BLSFixPer\_Depth\_1 & Transit depth at half the period returned by CETRA \\
BLSFixPer\_Depth\_2 & Transit depth at twice the period returned by CETRA \\
BLSFixPer\_Qtran\_0 & Fractional duration $q \equiv T_{\mathrm{dur}}/P$ of the primary signal \\
BLSFixPer\_Npointsintransit\_0 & Number of in-transit points for the primary signal \\
BLSFixPer\_Ntransits\_0 & Number of observed transits for the primary signal \\
BLSFixPer\_Ntransits\_1 & Number of observed transits at half the period returned by CETRA \\
BLSFixPer\_Ntransits\_2 & Number of observed transits at twice the period returned by CETRA \\
BLSFixPer\_Ntransits\_3 & Number of observed transits at three times the period returned by CETRA \\
BLSFixPer\_fraconenight\_0 & Fraction of in-transit points from a single transit (primary) \\
BLSFixPer\_fraconenight\_2 & Fraction of in-transit points from a single transit (at twice the period returned by CETRA) \\
BLS\_HarmAmp\_0 & Amplitude of the strongest harmonic signal \\
BLS\_HarmMean\_0 & Mean level of the strongest harmonic signal \\
BLS\_HarmDeltaChi2\_0 & $\Delta\chi^2$ improvement associated with harmonic variability \\
Teff & Stellar effective temperature \\
lum & Stellar luminosity \\
\enddata
\tablecomments{BLS-based parameters are derived from fixed-period BLS searches and their harmonics as well as a narrow period band BLS call. Stellar parameters provide host-star context during classification.}
\end{deluxetable*}
\end{center}
\FloatBarrier
\clearpage

\bibliographystyle{aasjournal}
\bibliography{references}

\end{document}